\documentclass[twocolumn]{aastex631}
\usepackage{booktabs}
\usepackage[flushleft]{threeparttable}
\DeclareRobustCommand{\ion}[2]{%
\relax\ifmmode
\ifx\testbx\f@series
{\mathbf{#1\,\mathsc{#2}}}\else
{\mathrm{#1\,\mathsc{#2}}}\fi
\else\textup{#1\,{\mdseries\textsc{#2}}}%
\fi}

\begin{document}

\title{PyEMILI: A New Generation Computer-aided Spectral Line Identifier}

\author[0009-0000-7976-7383]{Zhijun Tu}
\affiliation{CAS Key Laboratory of Optical Astronomy, National Astronomical Observatories, Chinese Academy of Sciences, 20A Datun Road, Beijing 100101, P.~R.\ China}
\affiliation{School of Astronomy and Space Sciences, University of Chinese Academy of Sciences, Beijing 100049, P.~R.\ China}

\author[0000-0003-1286-2743]{Xuan Fang}
\affiliation{CAS Key Laboratory of Optical Astronomy, National Astronomical Observatories, Chinese Academy of Sciences, 20A Datun Road, Beijing 100101, P.~R.\ China}
\affiliation{School of Astronomy and Space Sciences, University of Chinese Academy of Sciences, Beijing 100049, P.~R.\ China}
\affiliation{Xinjiang Astronomical Observatory, Chinese Academy of Sciences, 150 Science 1-Street, Urumqi, Xinjiang, 830011, P.~R.\ China}
\affiliation{Laboratory for Space Research, Faculty of Science, The University of Hong Kong, Pokfulam Road, Hong Kong, P.~R.\ China}

\author{Robert Williams}
\affiliation{Department of Astronomy \& Astrophysics, University of California, Santa Cruz, 1156 High Street, Santa Cruz, CA 95064, USA} 
\affiliation{Space Telescope Science Institute, 3700 San Martin Drive, Baltimore, MD 21218, USA} 

\author{Jifeng Liu}
\affiliation{CAS Key Laboratory of Optical Astronomy, National Astronomical Observatories, Chinese Academy of Sciences, 20A Datun Road, Beijing 100101, P.~R.\ China}
\affiliation{School of Astronomy and Space Sciences, University of Chinese Academy of Sciences, Beijing 100049, P.~R.\ China}
\affiliation{Institute for Frontiers in Astronomy and Astrophysics, Beijing Normal University, Beijing 102206, P.~R.\ China}
\affiliation{New Cornerstone Science Laboratory, National Astronomical Observatories, Chinese Academy of Sciences, Beijing 100101, P.~R.\ China}

\correspondingauthor{Xuan Fang}
\email{fangx@nao.cas.cn}

\begin{abstract}
Deep high-dispersion spectroscopy of Galactic photoionized gaseous nebulae, mainly planetary nebulae and \ion{H}{ii} regions, has revealed numerous emission lines.  As a key step of spectral analysis, identification of emission lines hitherto has mostly been done manually, which is a tedious task, given that each line needs to be carefully checked against huge volumes of atomic transition/spectroscopic database to reach a reliable assignment of identity.  Using {\sc Python}, we have developed a line-identification code PyEMILI, which is a significant improvement over the {\sc Fortran}-based package EMILI introduced $\sim$20 years ago.  In our new code PyEMILI, the major shortcomings in EMILI's line-identification technique have been amended.  Moreover, the atomic transition database utilized by PyEMILI was adopted from Atomic Line List v3.00b4 but greatly supplemented with theoretical transition data from the literature.  The effective recombination coefficients of the \ion{C}{ii}, \ion{O}{ii}, \ion{N}{ii} and \ion{Ne}{ii} nebular lines are collected from the literature to form a subset of the atomic transition database to aid identification of faint optical recombination lines in the spectra of PNe and \ion{H}{ii} regions.  PyEMILI is tested using the deep, high-dispersion spectra of two Galactic PNe, Hf\,2-2 and IC\,418, and gives better results of line identification than EMILI does.  We also ran PyEMILI on the optical spectrum of a late-type [WC11] star UVQS~J060819.93-715737.4 recently discovered in the Large Magellanic Cloud, and our results agree well with the previous manual identifications.  The new identifier PyEMILI is applicable to not only emission-line nebulae but also emission stars, such as Wolf-Rayet stars. 
\end{abstract}

\keywords{Gaseous nebulae (639), Planetary nebulae (1249), \ion{H}{ii} regions (694), Spectral line identification (2073), Atomic physics (2063), Spectroscopy (1558), Astronomy software (1855)}

\section{Introduction}
\label{sec1}

\subsection{Emission-line Nebulae and Identification of Emission Lines}
\label{subsec:background}

Planetary nebulae (PNe) and \ion{H}{ii} regions belong to the interstellar medium (ISM) in a galaxy, and are both photoionized nebulae that are well visible in the ultraviolet (UV) to optical wavebands.  PNe evolve from the low- to intermediate-mass ($\sim$0.8--8$M_{\odot}$) stars that cover a very broad range in main-sequence age, and are a short transition phase from the post-asymptotic giant branch (post-AGB) to the final white dwarf (WD) stage; the content of various heavy elements in PNe manifests the chemical environment of the ISM where their progenitor stars originally formed.  PNe are also the only ISM that exist in almost every part of a galaxy, from the bulge to the inner- and outer-disk and further to the halo regions; they are important tracers of local chemistry and stellar population.  \ion{H}{ii} regions are the interstellar ionized hydrogen associated with young (and relatively massive) stars, and are mostly concentrated in the galactic disc; they represent the current chemical environments of the host galaxy.  Deep spectroscopy of Galactic PNe and \ion{H}{ii} regions in the UV-optical wavelength region has revealed extremely rich emission lines, and thus they are both called emission-line nebulae.  However, the nomenclature ``emission-line nebulae'' has a more general meaning; it encompass various types of objects such as PNe, \ion{H}{ii} regions (including giant extragalactic \ion{H}{ii} regions), Wolf-Rayet bubbles, starburst galaxies, active galactic nuclei (quasars), and Herbig-Haro objects. 

The observed strength, wavelength and width of an emission line carry important information about the region where this line is emitted.  Emission lines are measured and routinely analyzed to derive the electron temperature ($T_{\rm e}$) and electron density ($N_{\rm e}$) of PNe and \ion{H}{ii} regions; this is called plasma diagnostics.  Based on the derived $T_{\rm e}$ and $N_{\rm e}$, ionic abundances can then be derived using the emission line fluxes relative to that of H$\beta$ 4861\,{\AA}.  Almost all the basic properties of PNe and \ion{H}{ii} regions are derived from the analysis of specific emission lines.  Hence it is apparent that correct identification of spectral lines is a crucial part of spectroscopic analysis.  However, over a century of spectroscopic observations of PNe and \ion{H}{ii} regions, only a small number of emission lines have been credibly identified manually.  High-resolution spectra of PNe can now be obtained through deep spectroscopy using ground-based telescopes, which reveals hundreds of emission lines in the optical wavelength region; identification of these emission lines can be a tedious task \citep[e.g.][]{2011MNRAS.415..181F,2012A&A...538A..54G,2018MNRAS.473.4476G}.  Besides, deep spectra of emission-line objects like supernova remnants, Wolf–Rayet (WR) stars and Herig-Haro (HH) objects, also become routinely available, which also need systematic line identifications \citep[e.g.][]{2020ApJ...888...54M}. 

Among all the emission-line objects, PNe exhibit the most complicated and numerous spectral lines, given that PNe have a central ionizing source (most usually a very hot white dwarf) and host a range of ionized, neutral and molecular material \citep[e.g.][]{OF2006}.  In the classical view, the ionization structure of a PN is governed by the distance of nebular material to the central ionizing star, and displays a stratified structure with higher ionization species such as He$^{2+}$ and O$^{2+}$ close to the central star, lower ionization species such as N$^{+}$ and O$^{+}$ in the outer region, and neutral and molecular species such as O$^{0}$ and H$_{2}$ in the outermost photo-dissociation region. 

Identification of spectral lines of PNe is a complicated decision process, in which various factors such as line intensity, presence or absence of other fine-structure components belonging to the same multiplet, line width and profile, etc., play important roles.  Experienced researchers identify spectral lines through a traditional and reliable approach: start with seeking for line identifications available in the literature for the objects with similar characteristics (excitation class, age, metallicity, etc.), then determine the identifications that are appropriate in the physical condition through wavelength match, comparison with the predicted line intensity, and presence or absence of other fine-structure transitions from the same multiplet (i.e.\ the multiplet check).  This traditional method, however, is extremely time-consuming and easy to introduce subjective biases. 

One technique of spectral line identification is to construct the synthetic spectra based on the existing atomic transition database, and then fit them to (or compare them against) the observed spectra.  This method can flexibly deal with line blending and wavelength errors very well.  However, the completeness and accuracy of the atomic database are the cornerstone of this method.  It's difficult to identify the faint lines that have no atomic transition data from the previous calculations.  The CHIANTI\footnote{https://www.chiantidatabase.org/} database aims to provide atomic data for solar and astrophysical environments to generate synthetic spectra of emission lines for analyses of spectroscopic observations of optically thin plasma \citep{CHIANTI}; it is being continually updated \citep{2023ApJS..268...52D}.

Machine learning techniques are becoming a focus of attention in astronomical research, as large-scale sky surveys have produced huge volumes of data.  A self-training spectral line identification code has been developed by \citet{2022JAP...132r3302T}.  Compared to the model-based approach, machine learning is more automated in providing relatively rigorous results of line identification.  However, the current disadvantage of machine learning is that it does not have a large number of training sets of deep spectra with a high confidence level of line identification.  Besides, the algorithm of this technique is highly dependent on training datasets; no physical parameter or interpretation ($T_\mathrm{e}$, $N_\mathrm{e}$, ionization structure, abundances, etc.) of the nebulae are involved in the process of identification.

The EMILI code developed 20\,yr ago offers relatively comprehensive line identifications for PNe and \ion{H}{ii} regions \citep[][]{2003sharpee}, with the objective of identifying numerous weak optical recombination lines (ORL) of heavy elements in high-dispersion spectra with high signal-to-noise (S/N) ratio.  Similar to CHIANTI, EMILI utilized a model-based technique.  The same logic as the traditional method is applied by EMILI, whereby the code first reads a list of measured emission lines and a large atomic transition database, and then ranks the candidate IDs based on wavelength agreement and predicted relative intensities, as well as on the presence/absence of other fine-structure component line(s) within the same multiplet as a given candidate ID.  The primary difference between EMILI and other line-identification techniques is that EMILI takes into account the cross-identification among the observed emission lines, while the comparison between the predicted and observed intensities is only one part of the line identification.  Therefore, EMILI could utilize a more complete atomic transition database to attempt line identifications even if some of the atomic data (e.g.\ transition probabilities) do not exist.

\subsection{The EMILI Code}
\label{subsec:emili}

Here we briefly describe the workflow of EMILI in identifying the emission lines of a PN.  Firstly, an \textit{Input Line List} with is compiled, with columns of (1) the measured wavelengths corrected to the rest frame, (2) the measured wavelength uncertainties (1$\sigma$), and (3) the extinction-corrected line fluxes with respect to $\mathrm{H\beta}$.  For each observed line in the list, EMILI draws from the atomic transition database the candidate IDs whose wavelengths are within 5$\sigma$ of the measured wavelength of this line.  EMILI utilizes the Atomic Line List v2.04\,\footnote{P.~A.~M.\ van Hoof. 1999, Atomic Line List v2.04, \url{http://www.pa.uky.edu/~peter/atomic/.}} database, which contains over 280,000 transitions covering a wavelength range of $\sim$3000--11,000\,{\AA}. 

Secondly, if some characteristic nebular emission lines -- usually those that are strong and easily identifiable -- are manually identified in the \textit{Matched Line List}, they will be used to construct a simple nebular model with the ionization and velocity structures based on the ionization potentials of the emitting ions and the internal radial velocities as measured from the observed nebular emission lines.  Otherwise, EMILI uses the default nebular model.  Thirdly, EMILI calculates the predicted template fluxes for all candidate IDs based on the nebular model.  Fourthly, for each candidate ID, EMILI searches in the \textit{Input Line List} for possible lines that belong to the same multiplet as the candidate ID.  Finally, EMILI ranks each candidate ID according to three criteria: (1) wavelength agreement, (2) comparison with the predicted template flux, and (3) the number of the multiplet members that are present in the \textit{Input Line List}.  Based on the three criteria, an identification index (\emph{IDI}) is defined to quantify the reliability of the candidate ID, 
\begin{equation} \label{eq1}
   IDI= W + F + M,
\end{equation}
where $W$, $F$, and $M$ are scores derived according to the wavelength agreement, comparison with the predicted template flux, and multiplet check, respectively, with lower scores better satisfied with the criteria.  All candidate IDs for a given emission line are then ranked as ``A'', ``B'', ``C'', ``D'', or ``None'' based on their $IDI$ values (e.g.\ IDs with the lowest $IDI$ will be ranked as ``A''). 

EMILI provides qualified line identifications on the premise that some characteristic emission lines (like those that are indicative of nebular ionization structures such as \ion{He}{i} $\lambda\lambda$4471,5876, \ion{He}{ii} $\lambda$4686 and highly ionized iron lines; see \citealt{2003sharpee}, Table\,1 therein) have been identified previously.  For the line identifications of Galactic PN IC\,418 and the Orion Nebula, the agreement between EMILI's results and those of the previous manual identifications is over 75\% and 85\%, respectively.  Here the percentage of agreement is defined with the number of emission lines whose manually assigned IDs are also ranked as ``A'' by EMILI.

\subsection{Limitations of EMILI}
\label{subsec:de_emili}

EMILI identifies reasonably well the majority of the collisionally excited lines (CELs, often referred to as forbidden lines given that most of these lines are forbidden transitions) detected in a nebular spectrum.  However, its identification of numerous faint optical recombination lines of heavy elements (such as C, N, O, and Ne) was unsatisfactory, and the number of reliable assignments to the observed emission lines is insufficient for nebular analysis.  Furthermore, the update of EMILI ceased since 2004.  Inadequate approximations, oversimplified nebular models, technical deficiencies, and some inherited bugs in EMILI result in limited capability of this code and hamper its further application in astrophysical research.  Moreover, EMILI was written in {\sc Fortran}~77, thus having poor compatibility on nowadays working stations.

The primary deficiencies in EMILI that need to be tackled with are:  (1) The Atomic Line List v2.04 utilized by the code is inadequate for the deep high-dispersion spectra of PNe and \ion{H}{ii} regions.  This atomic transition database does not include the \ion{He}{i} lines with the principal quantum number of the upper level larger than 15 ($n>$15) as well as the \ion{H}{i} lines with $n>$40.  For the relatively abundant heavy elements such as \ion{C}{ii}, \ion{N}{ii} and \ion{O}{ii}, transitions with $n>$5 (and probably the orbital angular momentum quantum number $l>$4) are not included in the Atomic Line List v2.04 either.  (2) The predicted template fluxes calculated by EMILI are mostly based on simple formulae and are prone to be unreliable, because the majority of transitions in the atomic database do not have references for transition probabilities or effective recombination coefficients.  The incorrect estimation of predicted template fluxes could affect the results of line identification.  (3) As for the multiplet check, EMILI only checks on those transitions that follow the pure LS-coupling scheme; while rigorous definition of a multiplet in the intermediate coupling schemes is difficult, especially for the atomic transitions with the upper-level angular momentum quantum number $l>3$, such as the 3d--4f transitions of the relatively abundant species \ion{O}{ii} and \ion{N}{ii}.  (4) A qualified result provided by EMILI always needs pre-identification of the characteristic lines (i.e.\ \textit{Matched Line List}), which complicates the process of line identification.

In this paper, we introduce a new {\sc Python}-based EMIssion-Line Identifier, PyEMILI.  Improvements in line-identification method of this new code over previous EMILI are elaborated in section \ref{sec2}.  In section~3, we provide a comparison of line identification results of PyEMILI and EMILI, using the same \textit{Input Line List} of Galactic PN IC\,418, which was identified by EMILI; we also identify the emission lines of another Galactic PN Hf\,2-2 as well as a late-type Wolf$-$Rayet [WC11] star UVQS J060819.93-715737.4 recently discovered in the Large Magellanic Cloud.  In section~4, we discuss the problems in developing PyEMILI and future improvement.

\begin{figure*}
\figurenum{1}
\centering 
\includegraphics[width=0.9\textwidth]{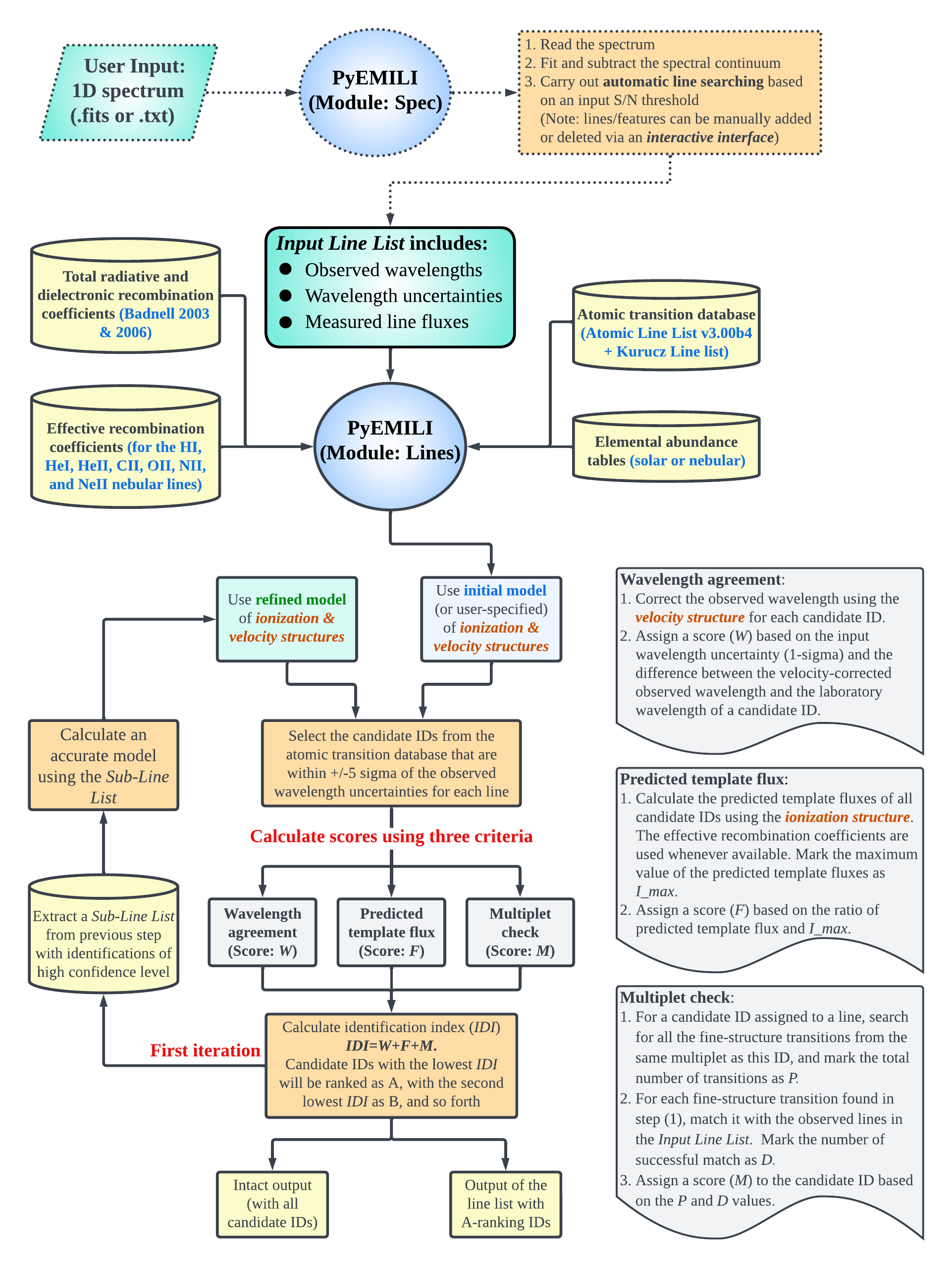}
\caption{Workflow chart of PyEMILI.  The three main criteria of PyEMILI's line identification are summarized separately in the three grey boxes in the bottom-right corner.} 
\label{flowchart}
\end{figure*}

\section{PyEMILI} 
\label{sec2}

\subsection{A Brief Introduction}
\label{pyemili:sec1}

PyEMILI is a new generation line-identification code developed on \textsc{Python\,3}, with a general purpose of spectral-line identification for high-quality, high-dispersion astrophysical spectroscopy with a broad wavelength coverage.  The workflow of PyEMILI is visually demonstrated in Figure~\ref{flowchart}, where the first few steps outlined by dotted lines are newly added subroutines, which are optional, to assist the users in searching the emission and/or absorption lines in an input spectrum as well as generating an \textit{Input Line List} (see section \ref{line_finding}).  Users can also provide their own \textit{Input Line List} to be read by PyEMILI.  The method of line identification of PyEMILI is generally similar to that of EMILI, i.e., the reliability of identification is assessed with three criteria to yield an $IDI$ for each candidate ID.  A summary of the criteria used in PyEMILI's line-identification for each component is presented in Table~\ref{tab:criteria}.  Compared to EMILI, PyEMILI has been improved and optimized in several key aspects, which are described in the following sections of this paper.

\subsection{The Energy-bin Model of PyEMILI} 
\label{pyemili:sec2}

In order to perform line identifications in physical conditions similar to the target (i.e.\ a PN or an H~{\sc ii} region), PyEMILI constructs an energy-bin model, which consists of five distinct energy regions/bins, each representing a range of ionization potential as defined by representative ions.  This is similar to the model method adopted by EMILI. Each energy bin is defined by two parameters that indicate the ionization and velocity structures.  The ionization structure parameter is used to estimate the ionic abundances for the calculation of predicted template flux, while the velocity structure parameter is intended to correct for different velocities of the ions with different ionization potentials. Determining the proper values of these two parameters in each energy bin requires a pre-identified line list, which is manually identified from the original \textit{Input Line List}. 

In EMILI, the pre-identified line list is called \textit{Matched Line List} \citep{2003sharpee}.  The absence of the \textit{Matched Line List} will severely affect the accuracy of the energy-bin model and consequently the accuracy of line identifications.  Thus, as mentioned above, EMILI's line identification relies strongly on the use of \textit{Matched Line List}.  However, unlike EMILI, PyEMILI does not require the \textit{Matched Line List} to refine the parameters of the ionization and velocity structures. Instead, PyEMILI first identifies spectral lines from the original \textit{Input Line List} using the energy-bin model with default values (or user-specified values), to produce a \textit{Sub-Line List}, which usually includes the strong spectral lines with rigorous identifications. 

This \textit{Sub-Line List} will be used to calculate the accurate parameter values of the ionization and velocity structures, which are then used to help re-identify (i.e.\ the first iteration; see Figure~\ref{flowchart}) the spectral lines for the target nebula.  The criteria for a line to be considered in the \textit{Sub-Line List} are (1) having only one candidate with ranking ``A'', 
and (2) the $IDI$ value of the ``A''-ranking candidate should be lower than the next ``B''-ranking candidate of the same line by at least 2.  By checking through the output list, the spectral lines with the identifications that fulfill the above two criteria are exported into the \textit{Sub-Line List}. 

In EMILI's \textit{Matched Line List}, all transitions are classified according to the next ionization energies of the parent ions, while in PyEMILI's \textit{Sub-Line List}, all the lines are classified into 5 energy bins according to the minimum ionization energies that are needed to produce the parent ions of the transitions.  For example, in PyEMILI, the parent ion of the \ion{O}{ii} recombination line, $\mathrm{O^{2+}}$, is assigned to the bin based on the second ionization energy $\sim$35.1~eV, while in EMILI, it's based on the third ionization energy $\sim$54.9~eV.  The classification method of PyEMILI describes the model more accurately after testing.  For special cases, \ion{H}{i}, \ion{He}{i}, and \ion{He}{ii} recombination lines are assigned to Bins~2, 3, and 4, respectively.   The energy ranges defined for each energy bin of the model are presented in Table~\ref{tab:bin}. 

The parameters of the ionization strucutre and the velocity structure in Table~\ref{tab:bin} are first given default/initial values for the first identification of the original \textit{Input Line List}.  Then the \textit{Sub-Line List}, which is generated according to the aforementioned two criteria, is used to refine the energy-bin model, which again is used for the second round of identification.  This second identification produces two files:  One is the complete output identifications (each line with all possible candidate IDs), and the other is the final line list (each line with the most probable candidate ID, with complete transition information presented).

\subsubsection{The Ionization Structure Parameter}
\label{ISP}

The ionization structure parameter is used to estimate the expected ionic abundances, which are then used to calculate the predicted template fluxes of the candidate IDs assigned by PyEMILI in line identifications.  In order to derive the ionic abundances, a table of total elemental abundances is previously required. 
There are two pre-set elemental abundance tables in PyEMILI -- the solar abundance table \citep{2009ARA&A..47..481A} and the typical nebular abundance table \citep{OF2006}, which the users can specify for PyEMILI. 

If an ion $\mathrm{X}^{i+}$ and its next-higher ionization stage $\mathrm{X}^{(i+1)+}$ are both located in the same energy bin, the abundance of the ion $\mathrm{X}^{i+}$ is defined as the total elemental abundance (of X) multiplied by the value of the ionization structure parameter for this energy bin.  If $\mathrm{X}^{i+}$ and $\mathrm{X}^{(i+1)+}$ belong to different energy bins, the ionic abundance of $\mathrm{X}^{i+}$ is the elemental abundance multiplied by the average value of the ionization parameter for these two energy bins where the ions $\mathrm{X}^{i+}$ and $\mathrm{X}^{(i+1)+}$ are separately located.  For example, under the parameter values of energy bins in Table~\ref{tab:bin}, the abundance of $\mathrm{N^{2+}}$ is the elemental abundance of nitrogen multiplied by (0.5$+$0.4)/2, an average of the parameters for Bin~2 and Bin~3, because the $\mathrm{N^{2+}}$ and $\mathrm{N^{3+}}$ are located in Bin~2 and Bin~3, respectively.  However, the abundance of $\mathrm{N^{3+}}$ is the elemental abundance of nitrogen multiplied by 0.4, as both the $\mathrm{N^{3+}}$ and $\mathrm{N^{4+}}$ are in Bin~3. 

Emission lines in the \textit{Sub-Line List} are used to refine and optimize the ionization structure parameter for each energy bin, so that more realistic values can be given for a PN, whose spectral lines are being identified.  Here we designate the values of the ionization structure parameter from Bin~1 to Bin~5 as $N_1$ to $N_5$.  $N_1$ is defined by the median value of the ratios of all emission lines within Bin~1, 
\begin{equation}
\label{eq:n1}
    N_1 = 0.01\times\text{Med}\left\{\frac{I_\mathrm{obs,Bin1}/I_\mathrm{H\beta}}{I_\mathrm{pre,Bin1}/I_{\mathrm{pre},\mathrm{H\beta}}}\right\},
\end{equation}
where $I_\mathrm{obs,Bin1}$ and $I_\mathrm{pre,Bin1}$ are the observed and predicted template fluxes for all lines in Bin~1.  The factor 0.01 in Eq~\ref{eq:n1} is used to normalize $N_1$ so that when the observed flux is consistent with the predicted template flux, it agrees with the initial value in Table~\ref{tab:bin}.  $N_5$ has the form 
\begin{equation}
\label{eq:n5}
    N_5 = I_\mathrm{obs,Bin5}/I_\mathrm{H\beta}+10^{-4},
\end{equation}
where $I_\mathrm{obs,Bin5}$ is the observed flux of the strongest emission line in Bin~5, given relatively few emission lines present in this energy bin in a typical nebula.  In a low-excitation PN, emission lines from Bin~5 can be very few or even absent; thus the second term $10^{-4}$ in Eq~\ref{eq:n5} is to ensure a minimum value of $N_5$ when no emission lines are observed from Bin~5. 

For $N_2$, $N_3$ and $N_4$, which are the parameter values of Bins~2, 3, and 4, we assume that the relative numbers of $\mathrm{H^+}$, $\mathrm{He^+}$ and $\mathrm{He^{2+}}$ are proportional to $N_2$, $N_3$ and $N_4$, respectively.  Therefore, we use the emissivities $\epsilon$ of $\mathrm{H\beta}$, \ion{He}{i} and \ion{He}{ii} lines. The emissivity $\epsilon$ (in units of ergs\,cm$^3$\,s$^{-1}$) is defined by the effective recombination coefficient (adopted from the literature, see section~\ref{ERC}) multiplied by photon energy
\begin{equation}
\label{eq:epsilon}
\epsilon_{\lambda} =\alpha_\mathrm{eff}(\lambda)\,\times\,h\nu.
\end{equation}
Consequently, $N_3$/$N_2$ is expressed as
\begin{equation}
\label{eq:n32}
    \frac{N_3}{N_2} = \frac{I_\mathrm{obs,\text{\ion{He}{i}}}/I_\mathrm{H\beta}}{\epsilon_\mathrm{\text{\ion{He}{i}}}/\epsilon_\mathrm{H\beta}},
\end{equation}
where the right side of Eq~\ref{eq:n32} is actually equivalent to the ionic abundance $\mathrm{He^+}$/$\mathrm{H^+}$. 
Similarly, the $N_4$/$N_2$ ratio is used for the \ion{He}{ii} line via 
\begin{equation}
\label{eq:n42}
    \frac{N_4}{N_2} = \frac{I_\mathrm{obs,\text{\ion{He}{ii}}}/I_\mathrm{H\beta}}{\epsilon_\mathrm{\text{\ion{He}{ii}}}/\epsilon_{{\rm H}\beta}}.
\end{equation}

For each \ion{He}{i} or \ion{He}{ii} emission line in the \textit{Sub-Line List} with available emissivity, PyEMILI calculates a ratio using Eq~\ref{eq:n32} or \ref{eq:n42}.  The final adopted $N_3$/$N_2$ and $N_4$/$N_2$ ratios are average values using all \ion{He}{i} and \ion{He}{ii} emission lines.  At last, we specify that a sum of the ionization parameters in all energy bins is unity,

\begin{equation}
\label{eq:sum}
    \sum_{i=1}^5 N_i=1. 
\end{equation}
Using Eqs.~\ref{eq:n1}--\ref{eq:sum}, we can explicitly derive the values for $N_1$, $N_2$, $N_3$, $N_4$ and $N_5$.  

\begin{deluxetable}{cccc}[tb]
\setlength{\tabcolsep}{0.5cm}
\tablecaption{Initial Values of Model Parameters for Different Energy Bins \label{tab:bin}}
\tablehead{\colhead{Bin} & \colhead{Energy range} & \colhead{ISP$^{a}$} & \colhead{VSP$^{b}$}\\ 
\colhead{} & \colhead{(eV)} & \colhead{} & \colhead{($\mathrm{km\,s^{-1}}$)} } 
\startdata
1 & 0--13.6 &  0.01 & 0\\
2 & 13.6--24.7 &  0.5 & 0\\
3 &  24.7--55 &  0.4 & 0\\
4 &  55--100 &  0.1 & 0\\
5 &  $>$100 & 0.001 & 0\\
\enddata
\tablecomments{\tablenotetext{a}{The initial value of the ionization structure parameter for each bin.}
\tablenotetext{b}{The initial value of the velocity structure parameter for each bin.}}
\end{deluxetable}

\subsubsection{The Velocity Structure Parameter}
\label{vel_structure}

In the standard structural model of a PN, it is usually assumed that the velocity structure is associated with the ionization structure: the high-ionization species are located close to the central star, while the lower-ionization species are in the outer nebular region; there is a velocity gradient from the central region to the outer region \citep{OF2006}.   Through deep, high-resolution spectroscopy of the Orion Nebula, \citet{2000ApJS..129..229B} found that the magnitude of difference between the observed wavelength (corrected for the systemic velocity of the nebula, the same meaning hereafter in this paper) and the laboratory wavelength of a nebular emission line is correlated with the ionization energy of the emitting ion.  The velocity structure parameter aims to correct for this correlation for the ions with different ionization energies.  The value of the velocity structure parameter in each energy bin is the average difference in the wavelengths (in $\mathrm{km\,s^{-1}}$) of all lines in the \textit{Sub-Line List} belonging to that bin.  If there are no lines in a bin, the value of the velocity structure parameter in that bin will adopt the value of the nearest bin with lines from the \textit{Sub-Line List}. 

In a typical PN, the velocity differences (between the observed wavelength and the laboratory wavelength) of the nebular emission lines with various ionization potentials generally exhibit small deviations from those derived using the \ion{H}{i} Balmer lines.  Consequently, during the initial phase of line identifications (utilizing the initial values of the velocity structure parameters), we assume that such velocity difference, which is thought to be associated with the ionization potential of the emitting ion, has no significant influence on the identification of strong emission lines.  These identifications rely more heavily on the predicted template fluxes of candidate IDs.  Therefore, we set an initial value of zero for the velocity structure parameters (Table\,\ref{tab:bin}) of all energy bins to simplify the initial modeling process and ensure logical consistency, avoiding potential biases introduced by artificial initial parameter values.

\begin{table}[t]
\caption{Criteria Used in Line Identification by PyEMILI\label{tab:criteria}}
\begin{tabular}{@{}ll@{}}
\toprule
\midrule
\multicolumn{2}{l}{\textbf{Wavelength Agreement}}                \\ 
\multicolumn{1}{l|}{$W$}   & Condition                    \\ \midrule
\multicolumn{1}{l|}{0}   & $\Delta\lambda$\footnote{$\Delta\lambda$ is the wavelength difference (in km\,s$^{-1}$) between the corrected observed wavelength and the laboratory wavelength of a candidate ID.} $\leq$1$\sigma$     \\
\multicolumn{1}{l|}{1}   & $1\sigma<\Delta\lambda\leq2\sigma$     \\  
\multicolumn{1}{l|}{2}   & $2\sigma<\Delta\lambda\leq3\sigma$     \\
\multicolumn{1}{l|}{3}   & $3\sigma<\Delta\lambda\leq4\sigma$     \\
\multicolumn{1}{l|}{4}   & $4\sigma<\Delta\lambda\leq5\sigma$     \\ \midrule
\toprule
\multicolumn{2}{l}{\textbf{Predicted Template Flux} }            \\ 
\multicolumn{1}{l|}{$F$}   & Condition                    \\ \midrule
\multicolumn{1}{l|}{0}   & $\geq 0.1\,I_{\rm max}$\,\footnote{$I_{\rm max}$ is the largest predicted template flux among all candidates.}\\
\multicolumn{1}{l|}{1}   & $\geq0.01\,I_{\rm max}$   \\
\multicolumn{1}{l|}{2}   & $\geq0.001\,I_{\rm max}$  \\
\multicolumn{1}{l|}{3}   & $\geq0.0001\,I_{\rm max}$ \\ \midrule
\toprule
\multicolumn{2}{l}{\textbf{Multiplet Check} }                  \\ 
\multicolumn{1}{l|}{$M$}   & Condition                    \\ \midrule
\multicolumn{1}{l|}{0}   & $P$\footnote{$P$ is the number of presumably observable fine-structure members within a multiplet.}=2, $D$\footnote{$D$ is the number of actually detected fine-structure members within a multiplet.}=2 or $D>$2 \\
\multicolumn{1}{l|}{1}   & $P =1, D=1$ or $P>2, D=2$                      \\
\multicolumn{1}{l|}{2}   & $P =0, D=0$ or $P>1, D=1$           \\
\multicolumn{1}{l|}{3}   & $P\leq2, D=0$           \\
\multicolumn{1}{l|}{4}   & $P >2,D=0$      \\ \midrule
\end{tabular}
\end{table}

\subsection{Atomic Transition Database}
\label{pyemili:sec3}

The completeness of an atomic transition database is critical for line identification by PyEMILI.  The most authoritative atomic transition database is probably the National Institute of Standards and Technology (NIST) Atomic Spectra Database (ASD)\footnote{\url{https://physics.nist.gov/PhysRefData/ASD/lines_form.html}}, where most wavelengths are laboratory values.  The transitions in NIST ASD are generally thought to be reliable, but still inadequate for identifications of emission lines in the deep high-dispersion spectra of nebulae -- in particular those obtained with the 8--10\,m class large optical-IR telescopes -- where new uncatalogued transitions might be detected.  Considering the wavelength ranges covered by most of the spectrographs on the ground-based and space-borne telescopes, we focus on the transitions in the wavelength region 1000--20,000\,{\AA}.  In this wavelength region, NIST ASD has compiled $\sim$200,000 transitions, which is much fewer than the Atomic Line List v2.04 database.  Many permitted transitions, in particular the numerous optical recombination lines of heavy elements detected in the deep spectra of PNe, are still missing from the NIST ASD, probably due to the difficulty in observing these lines on Earth.  Most of the emission lines excited under typical physical conditions of photoionized gaseous nebulae are impossible to detect in Earth labs, because of severe collisional de-excitation by particles \citep{2003ARA&A..41..517F}.  Consequently, we did not utilize the NIST ASD as the atomic transition database for PyEMILI.

The new Atomic Line List (hereafter AtLL) v3.00b4\footnote{\url{https://www.pa.uky.edu/~peter/newpage/}} 
developed by \citet{2018vanhoof} contains $\sim$900,000 atomic transitions from 1000\,{\AA} to 2.0\,$\mu$m, which is three times larger than the old version Atomic Line List v2.04 in number of transitions.  The transitions in AtLL v3.00b4 are not necessarily all observed, but both the upper and lower energy levels of every transition have been clearly catalogued.  Moreover, AtLL v3.00b4 is more rigorous in selecting energy levels, only including those with explicitly known quantum numbers (i.e.\ $n$, $l$, $s$, $j$).  The AtLL v3.00b4 is thus more comprehensive compared to the NIST ASD, and more suitable to be the atomic transition database of PyEMILI. 

Transition probability ($A_{\rm ji}$) is a key atomic data for line identifications, as it is used to predict the theoretical template flux, which is used to match the observed line flux.  The transitions and corresponding transition probabilities in AtLL v3.00b4 were collected and compiled from various sources.  Different sources introduce errors because of the methods of calculation.  Moreover, about two-thirds ($\sim$650,000) of the transitions included in AtLL v3.00b4 (in the wavelength region 1000--20,000\,{\AA}) miss transition probabilities.  We complement the AtLL v3.00b4 for the missing transition probabilities using the Kurucz Line Lists\footnote{\url{http://kurucz.harvard.edu/linelists/gfnew/gfall08oct17.dat}} \citep{2018ASPC..515...47K}.

The format of the Kurucz Line Lists is super-compressed because they were generated by a width-limited old machine called Line Printer; hence it needs to be careful in dealing with the format.  We matched the energies of the upper and lower levels of each of the transitions from the Kurucz Line Lists with the energies of the corresponding levels from AtLL v3.00b4 under a tolerance of $\Delta\,E\lesssim$0.1\,$\mathrm{cm^{-1}}$.  Moreover, for each transition, the ion and statistical weights of the upper and lower energy levels should be consistent in both databases.  As a result, $\sim$690,000 transition probabilities were updated and supplemented for the atomic transition database used by PyEMILI.  The final significantly expanded AtLL v3.00b4 includes huge numbers of transitions from the relatively abundant heavy elements in PNe, including \ion{O}{ii}, \ion{C}{ii}, \ion{N}{ii} and \ion{Ne}{ii}.

\subsection{The Criteria of Line Identification}
\label{IDcriteria} 

In Section\,\ref{subsec:emili}, we have introduced three main criteria in the emission-line identification by EMILI: wavelength agreement, the match of line flux, and the check for fine-structure transitions within the same.  Here we describe the three criteria we upgrade for PyEMILI.

\subsubsection{Wavelength Agreement}
\label{IDcriteria:1}

The criteria of wavelength agreement are based on the wavelength uncertainty (1\,$\sigma$), which is a key parameter in the \textit{Input Line List}, and the difference ($\Delta\lambda$) between the observed wavelength, which has been corrected for the systemic velocity of a PN as well as the internal velocity structure, and the laboratory wavelength of a candidate ID.  The wavelength uncertainty can be input as a single value applied to all emission lines to be identified, or specified individually and differently to each line.  Moreover, for each line with an observed wavelength $\lambda_{\rm obs}$, the upper and lower limits in wavelength uncertainty can be individually assigned.  Wavelength uncertainties can be largely affected by observations (e.g.\ spectral resolution) and data reduction (in particular, wavelength calibration), and may be significantly contributed by the uncertainty introduced by measurements.  The large measurement uncertainties, as well as the low resolution of a spectrum generally result in large wavelength uncertainties.  The score $W$ of wavelength agreement assigned to a candidate ID is defined according to the relation between $\Delta\lambda$ and $\sigma$ presented in Table~\ref{tab:criteria}.

In the identification of an observed emission line, the criteria of wavelength agreement using a smaller wavelength uncertainty can narrow the number of candidate IDs, so that the most likely assignment can be made.  However, the input wavelength uncertainties should not be too small, because the velocity structure parameters set for PyEMILI cannot be suitable for all emission lines in the \textit{Input Line List}, given that the actual velocity structure of the ionized gas in a PN can be always complex.  On the other hand, overly large wavelength uncertainties can result in too many candidate IDs, in particular the IDs with ranking ``A'', adding difficulty in scrutinization and correct assignment. 

Considering the facts mentioned above, as well as our experiences in the tests of PyEMILI, we choose typical values of wavelength uncertainty, 10--20\,$\mathrm{km\,s^{-1}}$ for all emission lines in the high-resolution ($R\geq$10,000--20,000) spectra, and 20--40\,$\mathrm{km\,s^{-1}}$ in the medium- to low-resolution  ($R<$10,000) spectra.

\subsubsection{Predicted Template Fluxes}
\label{IDcriteria:2}

Agreement in flux (i.e., comparison of the observed flux with the predicted flux of an emission line) is another criterion used in line identification.  In PyEMILI's identification of each emission line in the \textit{Input Line List}, the wavelengths of the items drawn from the atomic transition database as candidate IDs are within 5\,$\sigma$ measurement errors of this line.  For each candidate ID, PyEMILI calculates a predicted template flux using its transition probability and assuming a particular excitation mechanism appropriate to the physical condition of the PN (e.g., collisional excitation mainly for forbidden transitions, and radiative and/or dielectronic recombination for optical recombination lines; see the description hereafter in this section).  Only the transitions with predicted template fluxes larger than 10$^{-4}$ of the maximum value of the predicted fluxes ($I_{\rm max}$; Table~\ref{tab:criteria}) are retained as potential candidate IDs for an observed emission line. 

The grading of the scores $F$ of the predicted template fluxes is conducted on an order of magnitude scale.  In order to achieve a balance/consistency in the weighting of the three criteria used in line identification, it is necessary to assign each criterion the same range of scores.  Under this circumstance, the maximum value of the score $F$ of the predicted template flux should be 4, same as the criteria of wavelength agreement and multiplet check (see Table~\ref{tab:criteria}) as described in Sections~\ref{IDcriteria:1} and \ref{IDcriteria:3}, respectively.  The case of $F$ = 4 corresponds to the candidate IDs with very low values of predicted template fluxes, $\geq$10$^{-5}$\,$I_{\rm max}$.  The possibility of the candidate IDs with extremely low values of predicted fluxes is very low, given the detection limits of spectrographs.  On the other hand, the number of candidate IDs whose predicted template fluxes fall within the range of 10$^{-4}$--10$^{-5}$\,$I_{\rm max}$ is very high.  For the purpose of computational streamlining, we exclude the case of $F$ = 4.

Most of the emission lines detected in the optical spectra of PNe are produced by two mechanisms\footnote{Actually, Bowen fluorescence mechanism and charge transfer can also be at work in PNe, but are significant only for some particular transitions of a few ionic species \citep[e.g.][]{1993MNRAS.262..699L,1993MNRAS.261..465L}.}: collisional excitation and recombination.  In the optical spectrum of a PN, there are numerous nebular emission lines of \ion{C}{ii}, \ion{N}{ii}, \ion{O}{ii} and \ion{Ne}{ii} \citep[e.g.][]{2000liu,2003sharpee,2011MNRAS.415..181F,2013MNRAS.429.2791F}; these lines form through both radiative recombination (RR) and dielectronic recombination (DR).  In the typical physical conditions of PNe ($T_{\rm e}$\,$\sim$10$^{4}$\,K and $N_{\rm e}$\,$\sim$10$^{2}$--10$^{4}$\,cm$^{-3}$), the recombining ions mostly reside in the ground levels (i.e.\ fine-structure levels of the ground term; e.g., almost all the O$^{2+}$ ions of \ion{O}{ii} are in the $^{3}$P$_{J=0,1,2}$ levels), and the 3--3 (3s--3p and 3p--3d) and 3d--4f transitions are mainly contributed by direct radiative recombination. 

At very low electron temperatures ($T_{\rm e}\,\lesssim$1000\,K), radiate recombination dominates in the formation of the majority of the nebular recombination lines.  The radiative recombination coefficients decrease as $T_{\rm e}$ increase, because free electrons with lower velocities are easier to be captured (and recombined) by ions.  Indeed, previous deep spectroscopy has revealed that the ORLs of heavy elements are emitted from regions as low as $\lesssim$1000~K.  Above the temperature of 15,000--20,000\,K, contribution from dielectronic recombination is important; for many transitions the dielectronic recombination coefficient is greater than that of direct radiative recombination and becomes larger as $T_{\rm e}$ increases \citep[e.g.][]{1983A&A...126...75N}.  

In order to take into account different behaviors of radiative and dielectronic recombination in an extended range of electron temperature, we adopt a generic form of the predicted template flux, which is assumed to be proportional to the energy loss by recombination, in units of erg\,cm$^{-3}$\,s$^{-1}$ \citep{OF2006}, 
\begin{equation}
\label{recomb}
I_\mathrm{R} \propto L_{\rm R} = N_{\rm e}\,N_{i+1}\,C\,(\alpha_{\rm RR} + \alpha_{\rm DR})\,h\nu,
\end{equation}
where $N_{\rm e}$ is the electron density, $N_{i+1}$ is the number density of the recombining ion X$^{(i+1)+}$, and $\alpha_{\rm RR}$ and $\alpha_{\rm DR}$ are the total radiative recombination rate coefficient and the total dielectronic recombination rate coefficient, respectively, of the ground state of the ion X$^{i+}$.  The values of $\alpha_{\rm RR}$ and $\alpha_{\rm DR}$ were retrieved from \citet{2006badnellB}\footnote{\url{http://amdpp.phys.strath.ac.uk/tamoc/RR}} and \citet{2003badnell}\footnote{\url{http://amdpp.phys.strath.ac.uk/tamoc/DR}; a complete list of literatures is given on the website.}, respectively.  The total radiative recombination coefficient $\alpha_{\rm RR}$ has a relatively mild power-law dependence on electron temperature \citep[][Eq\,1 therein]{2006badnellB}.  The total dielectronic recombination rate coefficient, on the other hand, has a complex exponential dependence on electron temperature, and can be fitted as $\alpha_{\rm DR}$($T_{\rm e}$) = $T_\mathrm{e}^{-3/2}$\,$\sum_{i}\,c_{i}$\,exp($-E_{i}/T_\mathrm{e}$), where $c_{i}$ and $E_{i}$ (in units of eV or K) are fitting parameters given by the literature \citep[e.g.][Eq\,3 therein]{2003A&A...412..587Z}. 

The factor $C$ in Eq~\ref{recomb} is the branching ratio, which is proportional to the transition probability and converts the total recombination coefficients to effective recombination coefficients of lines.  Since the branching ratio of a specific level is difficult to estimate due to incomplete transition probabilities given in the literature, $C$ is approximated in the form 
\begin{equation}
\label{branch}
    C = \frac{A_{ji}}{\sum_{n=1}^{j-1}\sum_{m=n+1}^{j}A_{mn}},
\end{equation}
where $A_{ji}$ is transition probability from an upper level $j$ to lower level $i$.  The denominator in Eq~\ref{branch} is a sum of all the transition probabilities of the ion, whose upper level is not higher than $j$.  The transition probabilities of the enhanced atomic transition database used by PyEMILI are utilized to calculate $C$.  The term $C$\,($\alpha_{\rm RR}$ + $\alpha_{\rm DR}$) in Eq~\ref{recomb} is thus a general approximation of the effective recombination coefficient of an emission line.  We compare the approximated effective recombination coefficients with those calculated in the literature for \ion{O}{i} \citep{ERC_OI}, \ion{O}{ii} \citep{OIIcoe}, \ion{N}{ii} \citep{Fang_2011}, and \ion{N}{i}, \ion{C}{iii}, \ion{N}{iii}, \ion{N}{iv} and \ion{O}{iv} \citep{1991A&A...251..680P}; the vast majority of the differences are within a factor of $\sim$0.2 to 5. 

The transition probabilities in the enhanced atomic transition database are categorized into three distinct groups: the permitted (i.e.\ electric-dipole allowed) transitions, intercombination transitions, and forbidden transitions.  The transition probabilities within each group are arranged in ascending order according to their values, and the 25th percentile value is adopted for the transitions associated with the group of corresponding transition type that lacks transition probabilities.  As a result, we attribute (1) 10$^{4}$\,s$^{-1}$ to permitted transitions, (2) 10\,s$^{-1}$ to the intercombination transitions, and (3) 10$^{-5}$\,s$^{-1}$ to the magnetic-dipole, the electric-quadrupole, and even weaker transition types. 

For collisionally excited lines, we assume that their fluxes are proportional to the cooling rate, in units of erg\,cm$^{-3}$\,s$^{-1}$, as given by \citet[][see Chapter~3.5 therein]{OF2006},
\begin{equation} 
\label{collision}
I_\mathrm{C} \propto L_{\rm C} = D\,N_{\rm e}\,N_{1}\,q_{12}\,h\nu\,\frac{1}{1 + N_{\rm e}\,q_{21}/A_{21}},
\end{equation}
where $D$ is the dilution factor that can be adjusted so that the value of $I_{\rm C}$ can reach a better agreement with the observed strength of a collisionally excited line, $N_{1}$ is the number density of the ion X$^{\rm i+}$ populated in the lower level 1, and $q_{12}$ and $q_{21}$ are collisional excitation and collisional deexcitation coefficients (in units of cm$^{3}$\,s$^{-1}$), respectively.  The collisional deexcitation coefficient $q_{21}$ is expressed as 
\begin{equation}
q_{21} = \frac{8.63\times10^{-6}}{T^{1/2}_{\mathrm{e}}}\,\frac{\Upsilon(1,2)}{\omega_2},
\end{equation}
where $\Upsilon$(1,2) is velocity-averaged collision strength, whose value is between 0.1 and 10 for the major collisionally excited lines in the optical under the physical conditions (with a representative electron temperature of 10,000\,K) of photoionized gaseous nebulae \citep{OF2006}, and $A_{21}$ is spontaneous transition probability, for which we use the same rules as the recombination term.  For estimation of $I_{\rm C}$, we adopt a typical value of 1 for $\Upsilon$(1,2). 

The predicted template fluxes, either $I_{\rm R}$ or $I_{\rm C}$, are calculated based on the transition types of the lines.  We assume that the permitted transitions are mainly formed by recombination and the forbidden transitions are produced by collisional excitation; this assumption is reasonable and agree well with spectroscopic observations given the density and temperature conditions in the photoionized nebulae such as PNe and \ion{H}{ii} regions.  For the intercombination transitions (also called semi-forbidden transitions), the predicted template fluxes are a sum of the contributions from recombination and collisional excitation, but the flux contributed by the collision-excitation part is diluted by a factor of 100, which is an empirical value based on our investigation of transition probabilities.  For each observed emission line, PyEMILI calculates the predicted fluxes for all assigned candidate IDs, whose wavelengths are within 5$\sigma$ of that of the observed line, and retain those with predicted fluxes greater than a threshold value, which we set to be $10^{-4}$\,$I_{\rm max}$.  The criteria of flux comparison for PyEMILI's line identification is presented in Table\,\ref{tab:criteria}.

\subsubsection{The Multiplet Check} 
\label{IDcriteria:3}

In the identification of an emission line, besides wavelength match and flux comparison, it is also necessary to check through the whole observed spectrum (or the input line list) for the possible presence of other fine-structure lines originating from the same multiplet (a.k.a.\ multiplet members hereafter in this paper) as this line.  This helps to assess the robustness of line identifications.  Checking for multiplet members is particularly useful when a multiplet has more than three fine-structure components/transitions, as is the case for the relatively abundant heavy elements in PNe that have numerous, mostly faint optical recombination lines.  For example, the \ion{O}{ii} M1 2p$^{2}$3p $^{4}$D$^{\rm o}$ -- 2p$^{2}$3s $^{4}$P and \ion{N}{ii} M3 2p3p $^{3}$D -- 2p3s $^{3}$P$^{\rm o}$ multiplets, the strongest transitions of the two ions, have eight and six fine-structure transitions, respectively. 

Because of the very small energy differences between the fine-structure levels within a spectral term, all the fine-structure transitions belonging to the same multiplet have close wavelengths and are expected to have the same radial velocity.  Some multiplet members might even have comparable fluxes, although the line strengths of the fine-structure components within a multiplet can differ by several orders of magnitude (check, e.g., Figure~5 in \citealt{Fang_2011} and Figure~4 in \citealt{OIIcoe}, for the relative strengths of the fine-structure transitions within a multiplet).  For each candidate ID assigned to an observed emission line, PyEMILI searches through the atomic transition database for the other multiplet members and matches each of them against the observed lines in the \textit{Input Line List}.  In order to determine whether a given observed line being matched is a true/possible multiplet member of this candidate ID being tested, PyEMILI utilizes two judgment criteria.

The first criterion is the line-flux range: 
\begin{equation} \label{eq4}
0.1 < \frac{A_1g_1}{A_2g_2}/\frac{I_1}{I_2} < 10. 
\end{equation}
We designate $L1$ as the observed line that is being identified, and $T1$ as the candidate transition that is being tested corresponding to $L1$, and $L2$ and $T2$ as the similar meaning of another observed line and the candidate transition from the same multiplet as $T1$.  In Eq\,\ref{eq4}, $A_1$ and $g_1$ are the transition probability and statistic weight of $T1$, respectively, and $I_1$ is the observed (measured) intensity of line $L1$.  $A_2$, $g_2$ and $I_2$ have the same physical meaning for $T2$ and $L2$.  If either of these two transitions has no transition probability in the atomic transition database, PyEMILI will check if the observed intensity ratio of these two lines is within the range $0.1 < I_1/I_2 < 10$. 

The other criterion is to check the consistency in wavelength difference:  Assuming that both observed lines ($L1$ and $L2$) belong to the same multiplet, they are expected to origin from the same nebular region.  Consequently, the difference between the observed wavelengths (of $L1$ and $L2$), which are already corrected for the systemic velocity of a PN as well as the velocity structure within the PN (see Section\,2.2.2), and the corresponding laboratory wavelengths (of $T1$ and $T2$) should be within an uncertainty, i.e.\ 
\begin{equation} \label{eq5}
    |\Delta\lambda_1-\Delta\lambda_2|\leq\sigma,
\end{equation}
where $\Delta\lambda_1$ is the difference between the observed (and velocity corrected) wavelength of $L1$ and the laboratory wavelength of $T1$, and $\Delta\lambda_2$ has the same meaning for $L2$ and $T2$.  Here $\sigma$ is the wavelength uncertainty of $L1$. 

If the $T2$ transition fulfills the above two criteria, Eqs\,\ref{eq4} and \ref{eq5}, we call the $L2$ line the ``detected multiplet member'' (parameter $D$ in Table\,\ref{tab:criteria}) of the $T1$ transition (candidate ID of the $L1$ line) that is being tested. 

The multiplet check function in the EMILI code was designed for the pure $LS$-coupling scheme (i.e., only the transitions with the upper and lower levels with known $n$, $l$, $s$ and $j$ and labeled in the $LS$-coupling notation $^{2S+1}L^{P}_{J}$ can be recognized by EMILI) because of the difficulty in isolating those lines described in the $j$-$j$, $j$-$K$, and other coupling schemes from the atomic transition database.  This limitation has been well addressed in PyEMILI, which performs multiplet check for the spectral lines whose upper and lower energy levels are described in all coupling schemes (including $L$-$S$ coupling, $j$-$j$ coupling, and intermediate coupling), thanks to the parameterization of each atomic level that is reflected in the internal data file format\footnote{\url{https://linelist.pa.uky.edu/newpage/atll_format.html}} of AtLL v3.00b4. 

When reading the atomic transition database, PyEMILI follows the internal data format of AtLL, where each spectral term of an ion is assigned a unique integer.  When identifying an observed line, PyEMILI performs multiplet check for each candidate ID of the line.  For each candidate ID, PyEMILI selects from the atomic transition database the transitions that share the same spectral terms of the upper and lower energy levels as the candidate ID.  These selected transitions thus belong to the same multiplet as the candidate ID (i.e.\ they are multiplet members); the number of these transitions correspond to the parameter $P$ as introduced in Table\,\ref{tab:criteria}.  PyEMILI searches through the atomic transition database for the multiplet members of each candidate ID by matching the upper and lower spectral terms, which have been assigned unique integer numbers.  In this way, PyEMILI checks the multiplet members of all transitions regardless of coupling schemes.  It is thus capable of checking and identifying emission lines described in all major coupling schemes, and the reliability of line identification is secured. 

\smallskip

In order to further improve the accuracy of line identification, we incorporated {\it three extra criteria} into the multiplet check function of PyEMILI. 

(1) Table\,\ref{tab:criteria} lists the criteria of multiplet check, which we call the standard criteria.  We set a changeable flux threshold of $10^{-4}$ relative to H$\beta$ for PyEMILI to run the standard criteria of multiplet check.  In the special case of the lines with fluxes lower than $10^{-4}$ relative to $\mathrm{H\beta}$, the condition that ``$P>2, D=0$'' will be assigned a score $M=3$.  This extra criterion is to decrease the weight of multiplet check when the observed line fluxes are extremely low, to avoid the detection bias in extreme conditions where it is difficult to detect possible multiplet members in such low fluxes. 

(2) When the standard criteria of wavelength agreement and predicted template flux match of an observed emission line being identified by PyEMILI are both excellent (i.e.\ scores $W$=0 and $F$=0), but the score of multiplet check is assigned by PyEMILI to be $M$=3 or 4 (i.e.\ no other multiplet members of this observed emission line are found in the input line list; see the definition in Table\,\ref{tab:criteria}), we assign a score value of multiplet check to be $M$=2.  The corresponding total score $IDI$ is thus 2.  This artificial assignment is to reduced the weight of multiplet check as both the other two identification criteria have been assigned the best score by PyEMILI.  This extra criterion works well for the identification of high-dispersion emission-line spectra with low uncertainties in wavelength measurements, but will introduce new interference of candidates when the uncertainties in wavelengths are large. 

(3) For each candidate ID assigned to an observed emission line, PyEMILI will search through the output list to check whether the detected multiplet members of this observed line are ranked as ``A''.  If so, the detected multiplet members and the ranking of the candidate ID will be assigned a wedge symbol ``$\wedge$'' in front, as this candidate ID is also one of the best candidates of that observed line, regardless of whether it is ranked as ``A'' by PyEMILI.  The purpose of this criterion is to look for the candidates possibly blended with other strong emission lines.  Users can determine by themselves whether there is line-blending based on this annotation and line intensities. 
This extra criterion is used specifically to identify series of transitions that originate from high-$n$ (e.g.\ $n>$10 for \ion{He}{ii}), which are usually very faint in a spectrum.  For a faint emission line (labeled as $L1$) has been identified as a high-$n$ transition of \ion{He}{ii} (e.g.\ $n\rightarrow$2, labeled as $T1$) with ranking ``A'', PyEMILI will searches through its complete output list to check whether there is any candidate ID (with a corresponding observed line $L2$, of course) that belongs to the same series of the \ion{He}{ii} transition but with a lower $n$ (labeled as $T2$).  If the wavelength differences between the observed wavelengths of $L1$ and $L2$, and the laboratory wavelengths of $T1$ and $T2$ fullfills Eq\,\ref{eq5}, then PyEMILI labels in the specific output a wedge symbol ``$\wedge$'' in front of the $T2$ transition, regardless of whether $T2$ has been ranked ``A''; here ``$\wedge$'' has the same meaning as aforementioned.  In this way, PyEMILI searches and identifies all faint emission lines that are probably the high-$n$ transitions of  \ion{He}{ii} (and also \ion{H}{i}).

\begin{deluxetable}{llr}[t]
\tablecaption{References for the Effective Recombination Coefficients, $\alpha_{\rm eff}$($\lambda$), of ORLs Incorporated into the Atomic Transition Database of PyEMILI \label{ORLdata}}
\tablehead{\colhead{Atom/Ion} & \colhead{References for $\alpha_{\rm eff}$($\lambda$)} & \colhead{Comment}}  
\startdata
\ion{H}{i}   & \citet{HIcoe}   & Case~B \\
\ion{He}{i}  & \citet{HeIcoe}  & Case~B \\
\ion{He}{ii} & \citet{HIcoe}   & Case~B \\
\ion{C}{ii}  & \citet{CIIcoe}  & Case~B \\
\ion{N}{ii}  & \citet[][2013]{Fang_2011} & Case~B \\
\ion{O}{ii}  & \citet{OIIcoe}  & Case~B \\
\ion{Ne}{ii} & \citet{NeIIcoe} & Case~B \\
\enddata
\end{deluxetable}

\subsection{New Dataset of Atomic Transitions: Effective Recombination Coefficients} 
\label{ERC}

In the optical wavelength region, the absolute majority of the nebular lines detected in the deep, high-resolution spectrum of a PN are the ORLs emitted by the second-row heavy elements (mainly \ion{C}{ii}, \ion{N}{ii}, \ion{O}{ii} and \ion{Ne}{ii}); these ORLs are mostly faint, with fluxes $\lesssim$10$^{-4}$--10$^{-3}$ H$\beta$ at current detection limits of the instruments.  Identification of these ORLs with high accuracy is a key function of PyEMILI, and also relies on the effective recombination coefficients calculated specifically for the nebular lines. 

Along with the advance in observational techniques, in particular the advent of modern high-quantum-efficiency and large-format linear detectors, which have enabled detection of numerous faint ORLs of heavy-element ions in photoionized gaseous nebulae, the recombination theories of the relatively abundant ions, such as \ion{C}{ii}, \ion{N}{ii}, \ion{O}{ii} and \ion{Ne}{ii}, have been steadily improved since early 1980s \citep[e.g.][]{1981MNRAS.195P..27S,1983A&A...126...75N,1984A&AS...56..293N,1986A&AS...64..545N,1990ApJS...73..513E,1991A&A...251..680P,1994A&A...282..999S,1995MNRAS.272..369L,NeIIcoe,1999A&AS..137..157K,2002A&A...387.1135K,CIIcoe,Fang_2011,2013A&A...550C...2F,CIIdicoe,OIIcoe}.  The high-quality effective recombination coefficients, $\alpha_\mathrm{eff}$($\lambda$), have been widely used to reveal the physical conditions under which the ORLs of heavy-element ions arise, and to determine ionic and elemental abundances from them \citep[e.g.][]{McNabb_2012,2016MNRAS.461.2818M,2018MNRAS.473.4476G,2022MNRAS.510.5444G}.  Tools of plasma diagnostic for PNe and \ion{H}{ii} regions using the \ion{N}{ii} and \ion{O}{ii} ORLs have been well established based on the effective recombination coefficients \citep{McNabb_2012}. 

So far the effective recombination coefficients for the \ion{C}{ii}, \ion{N}{ii} and \ion{O}{ii} nebular lines are generally well calculated under the intermediate-coupling scheme \citep{CIIdicoe,Fang_2011,2013A&A...550C...2F,OIIcoe}; but the effective recombination coefficients of the \ion{Ne}{ii} nebular lines were only calculated in $L$-$S$ coupling \citep{NeIIcoe}, and are yet to be updated.  The effective recombination coefficients for the \ion{H}{i} (and \ion{He}{ii}) and \ion{He}{i} nebular lines, given the much simpler atomic structures of hydrogen and helium, have been calculated to very high levels of accuracy \citep{HIcoe,Benjamin_1999,Porter_2012,HeIcoe}. 

The datasets of these effective recombination coefficients $\alpha_{\rm eff}$($\lambda$) have been integrated into the atomic transition database used by PyEMILI to calculate the predicted template fluxes, which are used in the identification of nebular recombination lines detected in the deep spectra of PNe and \ion{H}{ii} regions.  The predicted template fluxes $I_{\rm pred}$($\lambda$) of the numerous faint nebular emission lines of \ion{C}{ii}, \ion{N}{ii}, \ion{O}{ii} and \ion{Ne}{ii}, normalized to $I$(H$\beta$) = 100, are calculated using the equation 
\begin{equation} \label{pred_flux}
I_{\rm pred}(\lambda) = \frac{I(\lambda)}{I({\rm H}\beta)} = \frac{\alpha_{\rm eff}(\lambda)}{\alpha_{\rm eff}({\rm H}\beta)} \frac{4861}{\lambda} \frac{{\rm X}^{2+}}{{\rm H}^{+}} \times 100 ,
\end{equation}
where X$^{2+}$ refers to C$^{2+}$, N$^{2+}$, O$^{2+}$ and Ne$^{2+}$.  $I_{\rm pred}$($\lambda$) is then used to compare with the observed line fluxes in line identification.  References for the effective recombination coefficients that have been collected and introduced into the database of PyEMILI are presented in Table\,\ref{ORLdata}.  It should be noted that all the effective recombination coefficients used by PyEMILI were calculated for Case~B, an assumption appropriate for the physical conditions in PNe and \ion{H}{ii} regions, and cover a wavelength range of 1000--20,000\,{\AA} for the nebular lines. 

For the effective recombination coefficients of the \ion{H}{i}, \ion{He}{i} and \ion{He}{ii} lines, calculations were carried out for the transitions with the upper-level principal quantum number $n\leq 25$.  It is necessary to mention that in the atomic transition database (as developed from AtLL v3.00b4 of \citealt{2018vanhoof}) used by PyEMILI, all the \ion{He}{i} transitions presented are $J$-resolved (i.e.\ presented as fine-structure transitions) and each of them has been assigned a unique ID, although the fine-structure components of \ion{He}{i} lines are seldom resolved in the actual spectroscopy.  On the other hand, the effective recombination coefficients of \ion{He}{i} lines so far calculated \citep[e.g.][]{Benjamin_1999,Porter_2012,HeIcoe} are all for the transitions between spectral terms (i.e.\ the calculations were all term-resolved, not $J$-resolved).  When PyEMILI calculates the predicted template flux of a fine-structure line of helium, it actually utilizes the term-resolved effective recombination coefficient of the He I transition; the predicted template flux thus calculated for a helium line is larger than that of individual fine-structure component.  

For those atomic lines ($J$-resolved transitions) without explicitly calculated effective recombination coefficients, their predicted template fluxes are calculated using Eq\,\ref{recomb}, which utilizes the total radiative recombination rate coefficient $\alpha_{\rm RR}$ and the total dielectronic recombination rate coefficient $\alpha_{\rm DRs}$ of the emitting ion, as well as the branching ratio $C$ (Eq\,\ref{branch}).  

Among all the heavy-element ions, the recombination theories of \ion{O}{ii} and \ion{N}{ii} are probably the most comprehensive, offering detailed calculations of the effective recombination coefficients for the individual fine-structure lines (transitions between $J$-resolved atomic levels) within a multiplet, at very broad ranges of electron temperature and electron density; in particular, the recombination calculations of these two ions are valid for electron temperatures down to an unprecedentedly low level, $<$200\,K \citep{2013A&A...550C...2F,OIIcoe}.  In contrast, the recombination calculation of the \ion{Ne}{ii} lines has seriously lagged behind, only carried out entirely in $L$-$S$ coupling \citep{NeIIcoe}; thus the $J$-resolved effective recombination coefficients for fine-structure transitions are unavailable for \ion{Ne}{ii}.  Nevertheless, the recombination calculations of the \ion{N}{ii}, \ion{O}{ii} and \ion{Ne}{ii} lines have all considered both radiative and dielectronic recombination processes. 

The dielectronic recombination lines of \ion{C}{ii} were calculated by \citet{CIIdicoe}, who considered autoionization (and near-threshold resonances) and utilized the $R$-matrix method in intermediate coupling.  However, these calculations are only for the dielectronic recombination process and therefore not suitable for the \ion{C}{ii} lines that are mainly produced through radiative recombination, e.g.\ the strongest transition of \ion{C}{ii}, M6 2s$^{2}$3d\,$^{2}$D -- 2s$^{2}$4f\,$^{2}$F$^{\rm o}$ $\lambda$4267 in the optical.  As a result, the current PyEMILI code makes use of the effective recombination coefficients for the \ion{C}{ii} lines calculated by \citet{CIIcoe}.

\begin{figure*}
\figurenum{2}
\centering 
\includegraphics[width=0.68\textwidth]{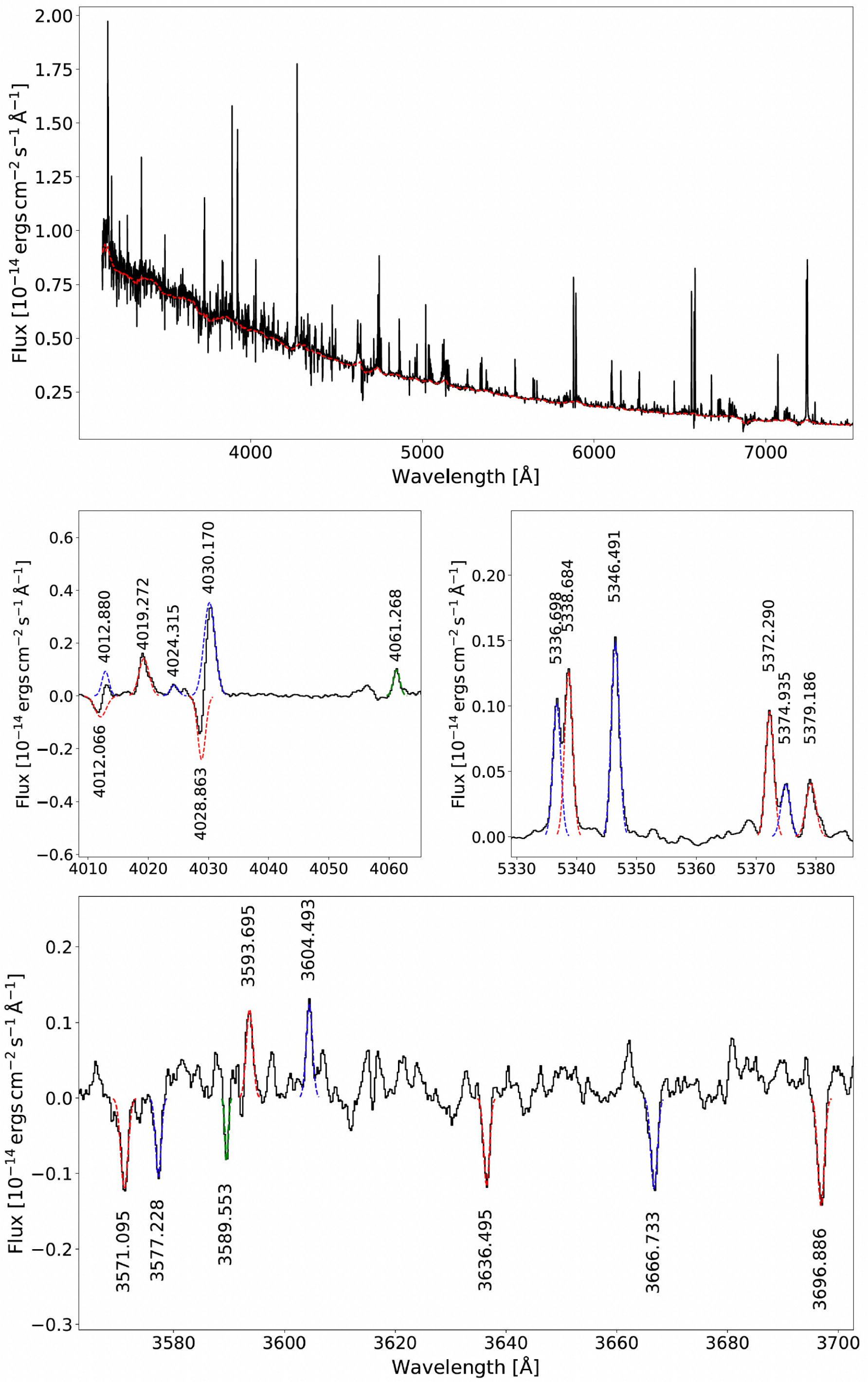}
\caption{\emph{Top}: Optical spectrum of an LMC [WC]-type star J0608.  The red curve is the result of automatic continuum fitting by a function in the module {\sc spec} of PyEMILI.  During the continuum fitting, the length of the moving window is set to be 50\,{\AA}, while the percentile parameter is set to be 30\%. 
\emph{Middle} and \emph{Bottom}: Examples showing automatic line searching in the spectrum (a) of J0608 by the PyEMILI module {\sc spec}.  The S/N ratio threshold of emission lines is set to be 7.  The continuum-subtracted spectrum is in black color, and the spectral lines automatically found by PyEMILI are fitted using the Gaussian profiles in red and blue; the two colors are used alternately to distinguish the boundaries of two lines in proximity.  The Gaussian profile in green in the \emph{Middle-left} and \emph{Bottom} panels is the line manually supplemented by the user through an interactive interface during line searching by the {\sc spec} module.  The \emph{Middle-left} panel demonstrate the fitting of P-Cygni profiles in the spectrum of J0608.  The \emph{Middle-right} panel shows the profile fit of partially blended emission lines.  The \emph{Bottom} panel demonstrates the {\sc spec} fitting of the absorption lines in the spectrum.  It should be mentioned here that the fitting of P-Cygni profiles is only a mathematically optimal solution.} 
\label{fig2}
\end{figure*}

To sum up, we have collected the effective recombination coefficients for the \ion{H}{i}, \ion{He}{ii}, \ion{C}{ii}, \ion{N}{ii}, \ion{O}{ii} and \ion{Ne}{ii} nebular lines from literature and arranged them into a separate atomic transition dataset to be used by PyEMILI; this dataset is used as an aid in the identification of numerous faint ORLs in the deep spectra of PNe and \ion{H}{ii} regions.  These effective recombination coefficients are used to calculate the theoretical fluxes (i.e.\ the predicted template fluxes) for identification of nebular lines.  During the multiplet check (see the description in Section\,2.4.3), when both emission lines ($L1$ and $L2$, as defined in Section\,2.4.3) have the effective recombination coefficients available, PyEMILI uses the following line flux criterion:
\begin{equation}
0.1 < \frac{\epsilon_1}{\epsilon_2}/\frac{I_1}{I_2} < 10,
\end{equation}
where $\epsilon_1$ and $\epsilon_2$ are the emissivities of transitions $T1$ and $T2$, calculated by Eq\,\ref{eq:epsilon} using the effective recombination coefficients.  More PyEMILI functions using the dataset of effective recombination coefficients are described in a separate paper (Z.\ Tu et al., in preparation).

\subsection{Applicability of PyEMILI} 

Like EMILI, PyEMILI was originally developed to identify the emission lines of photoionized gaseous nebulae, mainly PNe and \ion{H}{ii} regions.  Among the three criteria (wavelength agreement, predicted template fluxes, and multiplet check; see Section\,\ref{IDcriteria}) used by PyEMILI in line identifications, calculation/estimation of the predicted template fluxes is probably the most difficult, given that not all nebulae are known for their physical conditions (e.g.\ electron temperature and electron density) or ionic/elemental abundances, which are needed to predict the flux for an observed emission line.  For the emission-line nebulae, where the nebular lines are mainly formed through collisional excitation (i.e.\ CELs) and recombination (i.e.\ ORLs), PyEMILI calculates the predicted fluxes using the formulae as described in Section\,\ref{IDcriteria:2}. 

In order to make PyEMILI more versatile, we tried to enable it to identify the stellar absorption lines.  However, the formation mechanism of absorption lines in a stellar spectrum is totally different from that of emission lines in a nebular spectrum.  How the equivalent width (EW) of an absorption line is estimated in PyEMILI is briefly described in Subsection\,2.6.2.

\subsubsection{The Emission-line Spectra}

The calculation of predicted template fluxes (see Section\,\ref{IDcriteria:2}) may be applied in various emission-line objects other than PNe and \ion{H}{ii} regions, such as novae and supernova remnants (SNR).  The spectra of these nebulae exhibit some similarity.  As heated, ionized gas tends to emit the spectra with nebular emission lines in common, regardless of the exact mechanisms responsible for the high temperature and ionization required to produce the emission lines \citep{OF2006}.  The prescription of PyEMILI's line identification is also applicable to the emission spectra of Wolf-Rayet (WR) stars, where there are many broad emission lines of helium, carbon, and oxygen that are mostly formed through recombination, although with a more complex line formation process in reality \citep[e.g.][]{2007ARA&A..45..177C}. 

According to the observations \citep[e.g.][]{Gomez-Gonzalez2020,Gomez-Gonzalez_2021,2021MNRAS.508.2254R}, so far roughly 10\% of Galactic PNe have the central stars -- despite their low masses (typically $\sim$0.6\,$M_{\odot}$) -- exhibiting broad emission lines of helium, carbon and oxygen in their spectra, resembling the emission of classical Wolf-Rayet (WR) stars that evolve from massive progenitors.  In order to distinguish them from the classical WR stars, the low-mass PN central stars with broad emission features are denoted as [WR]; they are much older objects closely related to white dwarfs that are descended from evolved low-mass stars.  Theoretical predictions, as well as observations combined with the modeling of the post-AGB evolution, suggest that the low-mass [WR] stars can form through a final thermal pulse while the central stars of PNe descend along the white dwarf cooling track \citep{1983ApJ...264..605I,Guerrero_2018}; thus a low-mass [WR] star should always be associated with a PN.  In principal, PyEMILI can identify the emission lines in the spectrum of a low-mass [WR] star. 

If a PN surrounding a [WR] star is bright enough, nebular emission lines can be detected superimposed on the stellar spectrum.  However, most of the emission lines of [WR] stars originate from the stellar wind, which differs in nature from the photoionized nebulae.  The line-forming region in [WR] stars can have very high electron densities, causing collisional excitation to be suppressed. Besides, the stellar radiation field can significantly affect the stellar wind, and a high metallicity associated with it makes the dielectronic recombination process more intense \citep{2020ApJ...888...54M}. Thus, the electron temperature, electron density, dilution factor ($D$) for the CELs, and ionic abundances need to be modified based on the observed line intensities when identifying the lines of these objects.

\subsubsection{The Stellar Absorption-line Spectra} 

For an absorption line in a stellar spectrum, we calculate the theoretical EW of each candidate ID assigned by PyEMILI as its predicted template flux (in absorption) using the line-identification criteria described in Section\,\ref{IDcriteria:2}.  The theoretical calculation of EW must take into account the optical depth, because the value of the former has a dependence on the latter. 

To streamline the EW calculation for a transition, we consider a small optical depth (i.e.\ on the linear part of the curve-of-growth) to estimate the equivalent width $EW$ of the absorption line \citep[][Chapter\,3.4 therein]{1978ppim.book.....S}: 
\begin{equation} \label{eq:EW1}
    EW \propto \mathcal{N}_{\rm i} f_{\rm ik}\lambda^2,
\end{equation}
where $\mathcal{N}_{\rm i}$ is the column density of the ion that is populated in the lower quantum state/level ``i'' of the transition (and the upper quantum state/level is denoted as ``k''), which is defined as
\begin{equation}\label{eq:EW2}
    \mathcal{N}_{\rm i} = \int n_{\rm i}\,ds,
\end{equation}
integrated over the path along the line of sight.  Assuming that $\mathcal{N}_{\rm i}$ is proportional to the number density $n_{\rm i}$ of the ion populated in the lower state, and utilizing the Boltzmann equation that expresses the relationship between the number density of the ion in the lower state, $n_{\rm i}$, and the number density in the ground state, $n_{0}$, 
\begin{equation}\label{eq:EW3} 
    \frac{n_{\rm i}}{n_{0}} = \frac{g_{\rm i}}{g_{0}}\mathrm{e}^{-\Delta E/kT},
\end{equation}
where $\Delta E$ is the energy difference between the state i and the ground state (i.e.\ excitation energy of the state i), and $g_{\rm i}$ and $g_{0}$ are the statistical weights of the state i and the ground state, respectively. 

Given that in the physical conditions of stellar atmospheres, the ion mostly resides in the ground state, we assume that the number density of the ion $N$ is proportional to its number density populated in the ground state , $n_{0}$; thus we can derive the equation 
\begin{equation} \label{eq:EW4}
    \mathcal{N}_{\rm i} \propto N \frac{g_{\rm i}}{g_0} {\rm e}^{-\Delta E/kT}. 
\end{equation}
Here $g_{0}$ could be estimated as unity.  In Eq\,\ref{eq:EW1}, $f_{\rm ik}$ is oscillator strength, which is related to Einstein's spontaneous transition probability, $A_{\rm ki}$, through the formula 
\begin{equation} \label{eq:EW5}
    f_{\rm ik}=\frac{m_{\rm e}c}{8\pi^2{\rm e}^2} \frac{g_{\rm k}}{g_{\rm i}} \lambda^2 A_{\rm ki}.
\end{equation}
Finally, using Eqs~\ref{eq:EW1}, \ref{eq:EW4} and \ref{eq:EW5}, we obtain an approximation to estimate the equivalent width for an absorption line, 
\begin{equation} \label{eq:EW6}
    EW \propto N {\rm e}^{-\Delta E/kT} g_{\rm k} \lambda^4 A_{\rm ki}.
\end{equation}

The identification of absorption lines follows the procedure nearly identical to that of the identification of emission lines. For an absorption line to be identified, PyEMILI calculates the theoretical equivalent width for each candidate ID using Eq\,\ref{eq:EW6}.  The computed $EW$ serves as the predicted template flux for scoring and identification purposes.  Notably, during the multiplet check, the search for possible multiplet members (i.e.\ the other fine-structure components of the same multiplet as the line being identified) within the \textit{input line list} is carried out only among the absorption lines.  This criterion has been applied to emission lines (Section\,\ref{IDcriteria:3}).

\subsection{Automatic Line Searching}
\label{line_finding}

Manual measurements of the emission lines (their wavelengths, fluxes, widths, etc.) in a deep optical spectrum are a daunting task and very time-consuming.  In order to facilitate the generation of the list of massive emission lines from a deep spectrum of a photoionized gaseous nebula, we have designed for the PyEMILI code-suite a module, {\sc spec}, which has the functions to automate the processes of spectrum reading, continuum fitting, and spectral-line searching/fitting before line identification is carried out (Figure\,\ref{flowchart}). 

The continuum-fitting algorithm from {\sc alfa} \citep[Automated Line Fitting Algorithm;][]{ALFA}, which can automatically fit an emission-line spectrum of arbitrary wavelength coverage and resolution, was modified and adopted in the {\sc spec} subroutine. 
This fitting function works through a moving window with a user-specified width in wavelength; for an emission-line spectrum, the fluxes at all pixels within this window are used to calculate a 25th-percentile value that serves as the continuum at the midpoint (i.e.\ central wavelength) of the window.  This is similar to the median filter technique, but in our case (fitting the lower continuum in the emission line spectrum), the 25th percentile is more appropriate. 
The window then moves redward along the spectrum by one pixel, and the same computation is repeated.  This PyEMILI subroutine allows users to adjust the width of the moving window as well as the percentile value to accommodate various types of spectra, both emission and absorption.  For example, when dealing with a stellar spectrum dominated by absorption lines, a 75th percentile is usually used.  Moreover, the length of the moving window (in units of {\AA}) is determined based on the typical full-width at half-maximum (FWHM) of emission lines, with a wider window adopted for larger FWHMs. 

Typically this approach of continuum fitting is utilized through a moving window with a fixed length.  At the beginning and the ending points (pixels) of a spectrum, where the window length is half of the full length, no continuum flux can be calculated.  To maintain the integrity of the spectrum, the width of the moving window is adjusted at both boundaries by progressively increasing (when starting at the blue beginning) or reducing (when reaching the red end) its length to calculate the continuum.  Specifically, we consider the example of a moving window with 101 data points shifting redward along a spectrum. 

When the right-edge of the moving window reaches the red end of a spectrum, the central wavelength of the window, where the local continuum is calculated, is 50 data points/pixels away from the red end of the spectrum.  As the window continues to shift redward by $n$ data points, the distance from the red end of the spectrum to the wavelength point at which the continuum is calculated will be reduced to ($50-n$) data points.  The left-edge of the moving window remains progressively shifted redward, so that the overall window length is reduced to ($101-n$) data points.  Consequently, only ($101-n$) flux values are used to compute the local continuum.  A similar method is applied to the case where the central position of the moving window starts from the blue beginning of the spectrum.  This technique ensures the complete preservation of the original spectrum, though it may lead to larger continuum errors on both ends of the spectrum.

After the continuum fitting of an observed spectrum as described above, continuum is then subtracted to obtain a pure emission/absorption line spectrum, where we then carry out statistics of the S/N ratio also through a moving window with a user-defined width in {\AA}.  At each wavelength region within the window, PyEMILI determines the S/N ratio as a threshold to search for potential spectral lines, which are easily found in regions with high S/N.  The initial step involves the computation of the standard deviation of noise for the wavelength range within the window.  The sigma-clipping algorithm is used for this purpose, including the following two steps: 

(1) Within a window of the continuum-subtracted spectrum, the median value ($\mu$) of all fluxes and the standard deviation ($\sigma$) for the window are calculated.  The data points (i.e.\ pixels) with fluxes falling outside the range of $\mu\pm3\sigma$ are removed. 

(2) Using the remaining data points, we repeat step (1) until a predetermined stopping criterion is met, either a user-specified number of iterations or when the new standard deviation ($\sigma_{\rm new}$) falls within a user-defined tolerance of the previous iteration value, such as $(\sigma_{\rm old} - \sigma_{\rm new})/\sigma_{\rm new} < 1\%$. 

The final $\sigma$ obtained through this process is the standard deviation of the noise level of the spectrum covered in the window.  The window is then shifted to the longer wavelength by one window length, to calculate the standard deviation of the noise level for the subsequent wavelength region.

\smallskip

Following the computation of the sigma-clipped standard deviations of the wavelength region within the moving window, PyEMILI proceeds to determine the S/N ratio therein and find the local maxima (peaks) that exceed the user-specified threshold of S/N ratio.  The peak-finding algorithm used here is adopted from the \textsc{Scipy.signal.find\_peaks}\footnote{\url{https://docs.scipy.org/doc/scipy/reference/generated/scipy.signal.find_peaks.html}} function of the open-source software {\sc SciPy}. Additionally, PyEMILI utilizes the parameter $prominence$ (refer to the documentation of \textsc{Scipy.signal.find\_peaks}) to determine potential line blending in the spectrum.  When multiple peaks exist within a wavelength region being fitted, the $prominence$ parameter is used to determine whether the sub-peaks should be treated as independent spectral lines.  If more than one peak is identified in a specified region, PyEMILI will fit this wavelength region with a multi-Gaussian function, with the number of Gaussian profiles depending on the number of peaks.  In contrast, a single Gaussian profile is used when only one peak is found locally. 

Although very practical, there is limitation in automatic line searching due to different data quality and spectral resolution, as well as the distraction by the telluric emission/absorption and ghosts introduced by instrumental deficiency.  Therefore, in order to facilitate improving the performance of automatic line searching, we developed an optional function of interactive interface for PyEMILI using the {\sc Python}-based code {\sc matplotlib}\footnote{\url{https://matplotlib.org/}} to create visualizations in line searching.  The interactive interface can be activated upon execution of the code, which allows users to manually make real-time modifications during line searching on the spectrum being displayed. 

The interactive interface in PyEMILI enables users to zoom in, zoom out, and pan across the spectrum for detailed examination and scrutinisation of spectral lines.  The exact operation of this function generally follows standard manners, like those in {\sc iraf}.  For a spectral line that may have been missed by the code, users can mark it by first placing the cursor on the left side (i.e.\ left boundary at the continuum level) of this line, followed by clicking a designated key, and then placing the cursor on its right side, followed by anther click of the same key.  The program then performs the Gaussian-profile fit to the spectral data within the wavelength range specified by the cursor.  If the Gaussian-profile fit is successful, this line will be added to the list of detected emission lines, which will be read by PyEMILI as input. 

For a ``fake'' feature (usually very faint, not a real spectral line but noise or the residual of sky emission) that has been mistakenly marked as an emission line by PyEMILI, users can place the cursor on this feature and remove it from the list of detected lines.  This interactive approach of automatic line searching ensures that the spectral line list can be refined and corrected, improving the accuracy of the results.  Figure\,\ref{fig2} demonstrates the examples of automatic line searching of PyEMILI.

\subsection{Run Time of PyEMILI}
\label{run_time} 

The run time of the PyEMILI main code in identifying a list of 1000 lines, currently the maximum number of emission lines detected in the deepest spectrum of a bright Galactic PN in the wavelength range $\sim$3040--11,000\,{\AA} obtained with a 8--10\,m aperture large telescope, is typically within one minute, when on a standard processor (with a baseline speed 2.5\,GHz).  However, we need to specifically mention that, as some functions in the PyEMILI code utilize the {\sc Python} compiler {\sc Numba} to accelerate computation, the initial run on a device may require a few extra minutes for caching. 
When using PyEMILI’s line-searching module in line identification, the time needed to display the interactive interface and line-finding results is also typically within one minute.  However, extra time is needed in visual inspection, addition, and removal of lines/features in via an interactive window, and varies by individual users.

\begin{deluxetable*}{rllcrllrrrrcrr}
\rotate
\tabletypesize{\scriptsize}
\tablecaption{A Sample of PyEMILI Output Showing the Top 15 Candidates IDs of An Emission Line Observed at 3918.93\,{\AA} in IC\,418 \label{tab:exam_ic418}}
\tablehead{\colhead{(1)} & \colhead{(2)} & \colhead{(3)} & \colhead{(4)} & \colhead{(5)} & \colhead{(6)} & \colhead{(7)} & \colhead{(8)} & \colhead{(9)} & \colhead{(10)} & \colhead{(11)} & \colhead{(12)} & \colhead{(13)} & \colhead{(14)} 
 } 
\startdata
$+$3918.921 & 3918.97   &  \ion{C}{ii}    &   $\mathrm{{{^2}P^o}\text{--}{{^2}S}         }$ &  $\sim$4.78E$-$04 &  1/1  & 1  &  \string^A &  $-$3.57 &\string^3920.68 & $-$3.50    & $\mathrm{2s^2.3p\text{--}2s^2.4s                       }$  &    2 & 2 \\
$+$3918.937 & 3919.00   &  \ion{N}{ii}    &   $\mathrm{{{^1}P}\text{--}{{^1}P^o}         }$ &  $\sim$6.28E$-$06 &  0/0  & 3  &    B &  $-$4.84 &          &         & $\mathrm{2s^2.2p.({{^2}P^o}).3p\text{--}2s^2.2p.({{^2}P^o}).3d     }$  &    3 & 3 \\
$+$3918.881 & 3918.80     &  \ion{S}{i}     &   $\mathrm{{^3P}\text{--}{{^3}D^o}         }$ &        3.71E$-$07 &  0/0  & 5  &    C &  6.22 &          &         & $\mathrm{3s^2.3p^3.({^4S^o}).4p\text{--}3s^2.3p^3.({^2D^o}).4d   }$  &    3 & 5 \\
 3918.937 &  3918.80     &  \ion{C}{iii}   &   $\mathrm{{^3P^o}\text{--}{{^3}D}         }$ &         8.74E$-$06 &  2/0  & 5  &    C &  10.53 &          &         & $\mathrm{2p.3d\text{--}2s.7d                         }$  &    3 & 5 \\
 3918.881 &  3918.74  &  [\ion{Fe}{ii}] &   $\mathrm{a{{^4}D}\text{--}c{{^2}G}         }$ &          7.45E$-$06 &  2/0  & 5  &    C &  10.74 &          &         &  $\mathrm{3d^6.({^5D}).4s\text{--}3d^6.({^1G^2}).4s          }$  &    8 & 10 \\
 3918.881 &  3919.10     &  \ion{S}{i}     &   $\mathrm{{^3P}\text{--}{{^3}D^o}         }$ &         2.98E$-$07 &  0/0  & 6  &    D &  $-$16.7 &          &         & $\mathrm{3s^2.3p^3.({^4S^o}).4p\text{--}3s^2.3p^3.({^2D^o}).4d   }$  &    1 & 3 \\
 3918.955 & 3918.80     &  [\ion{Mn}{vi}] &   $\mathrm{{^3F}\text{--}{{^1}G}           }$ &          7.07E$-$07 &  2/0  & 6  &    D &  11.92 &          &         &  $\mathrm{3d^2\text{--}3d^2                           }$  &    5 & 9 \\
 3918.937 &  3919.27   &  \ion{O}{ii}    &   $\mathrm{{^2D}\text{--}{{^2}P^o}           }$ &   $\sim$2.75E$-$05 &  2/0  & 6  &    D &  $-$25.6 &          &         & $\mathrm{2s^2.2p^2.({^1D}).3s\text{--}2s^2.2p^2.({^1D}).3p     }$  &    4 & 2 \\
 3918.881 &  3918.42  &  \ion{Fe}{i}    &   $\mathrm{b{{^3}P}\text{--}x{{^3}P^o}       }$ &          8.34E$-$07 &  5/1  & 7  &      &  35.6 & 3925.94 & 41.9    & $\mathrm{3d^7.({^4P}).4s\text{--}3d^7.({^4P}).4p             }$  &    3 & 1 \\
 3918.881 &  3918.78  &  \ion{Fe}{i}    &   $\mathrm{c{{^3}F}\text{--}{{^{7/2}}[3/2]^o}   }$ &       5.25E$-$09 &  1/0  & 7  &      &  7.86 &          &         &  $\mathrm{3d^8\text{--}3d^7.({{^4}F\left \langle7/2\right \rangle}).4f              }$  &    5 & 5 \\
 3918.881 &  3919.30     &  \ion{S}{i}     &   $\mathrm{{^3P}\text{--}{{^3}D^o}         }$ &         5.97E$-$07 &  0/0  & 7  &      &  $-$32.0 &          &         & $\mathrm{3s^2.3p^3.({^4S^o}).4p\text{--}3s^2.3p^3.({^2D^o}).4d   }$  &    5 & 7 \\
 3918.881 &  3919.07  &  \ion{Fe}{i}    &   $\mathrm{b{{^3}G}\text{--}v{{^3}G^o}       }$ &          6.90E$-$8 &  5/0  & 8  &      &  $-$14.1 &          &         & $\mathrm{3d^6.4s^2\text{--}3d^7.({^2G}).4p                 }$  &    9 & 9 \\
$+$3918.881 &  3919.15   &  \ion{Cr}{i}    &   $\mathrm{a{{^5}D}\text{--}z{{^5}D^o}       }$ &         2.33E$-$06 &  3/0  & 8  &      &  $-$20.9 &          &         & $\mathrm{3d^4.4s^2\text{--}3d^4.({^5D}).4s.4p.({^3P^o})        }$  &    9 & 9 \\
 3918.881 &  3919.15   &  \ion{Cr}{i}    &   $\mathrm{b{{^5}D}\text{--}t{{^5}P^o}       }$ &         1.85E$-$07 &  1/0  & 9  &      &  $-$20.5 &          &         & $\mathrm{3d^5.({^4D}).4s\text{--}3d^5.({^4D}).4p             }$  &    7 & 7 \\
 3918.921 &  3918.52  &  [\ion{Mn}{iii}]    &   $\mathrm{a{{^2}H}\text{--}b{{^4}D}       }$ &  $\ast$1.21E$-$08 &  2/0  & 10  &      &  30.7 &          &         & $\mathrm{3d^5.4s^2\text{--}3d^4.({^5D}).4s                 }$  &    12 & 8 
\enddata
\tablecomments{
The nebular emission line in IC\,418 has an observed wavelength at 3918.93\,{\AA} (already corrected for the radial velocity of IC\,418 at $+$68.6\,km\,s$^{-1}$, \citealt{2003sharpee}), with an extinction-corrected flux of 1.07$\times$10$^{-3}$ relative to H$\beta$.  SNR$\sim$226; FWHM$\sim$18.7\,km\,s$^{-1}$. \\
\smallskip\\
Column (1): Wavelength of a candidate ID corrected using the velocity-structure parameter as defined in Section\,\ref{vel_structure}.  The ``+'' notation in front of the wavelength means that the velocity difference in Column (9) is less than 1$\sigma$. \\
Column (2): Laboratory wavelength of a candidate ID.  An asterisk ``$\ast$'' in front of the laboratory wavelength means it is a weighted-average wavelength if all the fine-structure transitions (including this candidate ID) within a multiplet have very close wavelengths, e.g.\ with differences smaller than the instrumental resolution.  No asterisk in this example, because it is not applicable to this case. \\
Column (3): The emitting ion. \\
Column (4): The lower and upper spectral terms. \\
Column (5): The predicted template flux relative to H$\beta$.   The value with a tilde ``$\sim$'' in front is calculated using the effective recombination coefficient (Section\,\ref{ERC}), while that with an asterisk ``$\ast$'' in front indicates the transition has no transition probability in our atomic transition database and its predicted template flux is estimated using the methods in Section\,\ref{IDcriteria:2}. \\
Column (6): The number of associated multiplet members that are expected to be observed, versus the number of transitions actually detected. \\
Column (7): The identification index (\emph{IDI}, see definition in Eq\,\ref{eq1}) of the candidate ID. \\
Column (8): Ranking of the candidate ID assigned by PyEMILI.  The wedge symbol ``$\wedge$'' in front of a ranking means at least one detected multiplet member has been ranked as ``A'' for the corresponding observed line.  In this case, the candidate ID is treated as one of the best identifications regardless of whether it has been ranked ``A''. \\
Column (9): Velocity difference (in km\,s$^{-1}$) between the wavelengths in Column (1) and Column (2). \\
Columns (10) and (11): The associated multiplet members (i.e.\ fine-structure components belonging to the same multiplet) detected from the \textit{Input Line List}, and their velocity difference, respectively.  Up to three multiplet line are displayed in this output.  The wedge symbol ``$\wedge$'' in front of the wavelength means this multiplet member is assigned ``A''-ranking in the corresponding observed line. \\
Column (12): Electron configurations of the lower and upper states. \\
Columns (13) and (14): Statistical weights of the lower and upper states, respectively. \\
\smallskip\\
(This table is published in its entirety online only in the machine-readable format, or accessible via the project website as presented in Section\,2.1) \\
(Readers can also contact the authors for a simplified version (806 lines) of PyEMILI's output for IC\,418) 
} 
\end{deluxetable*}

\begin{deluxetable}{crrrr}
\tablecaption{Number Statistics of Emission-line Identifications for IC\,418: EMILI versus PyEMILI\label{tab:ic418}} 
\setlength{\tabcolsep}{0.35cm}
\tablehead{\colhead{Rank} & \multicolumn{2}{c}{\underline{~~~~~~EMILI~~~~~~}} & \multicolumn{2}{c}{\underline{~~~~~PyEMILI~~~~~}} \\ 
\colhead{} & \colhead{(Number)} & \colhead{(\%)} & \colhead{(Number)} & \colhead{(\%)} } 
\startdata
A     &  524   &  75.3    &  636   &  91.3 \\
B     &   50   &   7.2    &   22   &   3.2 \\
C     &   19   &   2.7    &   18   &   2.6 \\
D     &   14   &   2.0    &   10   &   1.4 \\
None  &   89   &  12.8    &   10   &   1.4 \\
\enddata
\tablecomments{
A total of 696 emission lines were manually identified by \citet{2003sharpee} from the spectrum of IC\,418. \\
\smallskip\\ 
``None'' indicates the line identification is lower than D.} 
\end{deluxetable}

\section{Test of PyEMILI on Emission-Line Objects}
\label{sec3}

As stated above, PyEMILI has been designed to identify not only the numerous nebular emission lines of photoionized gaseous nebulae (mainly PNe and \ion{H}{ii} regions), but also the emission lines in WR stars as well as 
absorption lines in stellar spectra.  In this section, in order to test the overall performance -- in particular the reliability of line identification -- of PyEMILI, we carry out identification of the emission lines detected in the deep spectra of two Galactic PNe, Hf\,2-2 and IC\,418, and a late-type [WC11] star UVQS~J060819.93-715737.4 recently discovered in the Large Magellanic Cloud.  The spectra of the three objects used for PyEMILI test cover the whole optical wavelength range (from $\sim$3200--3500\,{AA} in the blue to $\sim$1\,$\mu$m in the red), and were obtained through medium- to high-dispersion ($R\sim$4100--30,000) spectroscopy with the echelle spectrographs on the 4--8\,m class optical telescopes.

\subsection{IC\,418}
\label{ic418}

The Galactic PN IC\,418 is the first object with a published line list systematically identified by EMILI \citep{2003sharpee}; thus it is an ideal target used to test PyEMILI and compare the overall performance of the two codes in line identification.  The Echelle spectroscopy of IC\,418 in the wavelength range $\sim$3500--9865\,{\AA}, with a spectral resolution of 9\,km\,s$^{-1}$ ($R\sim$33,000), was obtained using the CITO 4\,m Blanco telescope; reduction of the echelle data as well as the level of accuracy are described in \citep{2000ApJS..129..229B,2004sharpee}.  In total 807 emission lines were detected in the echelle spectra of IC\,418, with line fluxes measured to levels as faint as $\leq$10$^{-5}$ that of H$\beta$ \citep{2004sharpee}. 

We use the emission line list of IC\,418 compiled by \citet{2003sharpee} as the \textit{Input Line List} for PyEMILI.  In the line identifications of IC\,418 made by EMILI, the wavelength uncertainties of the emission lines as measured from the high-dispersion echelle spectroscopy of IC\,418 are set within the range of $\sim$5--10\,km\,s$^{-1}$.  While for the PyEMILI run, we set a uniform wavelength uncertainty of 10\,km\,s$^{-1}$ to all the lines, thus including more possible candidate IDs assigned to an emission line being identified.  This adjustment in the input wavelength uncertainty will have little effect on the identification of strong lines, but will improve the identification of weak lines, particularly those with larger wavelength uncertainties. 

As an example of the results, Table\,\ref{tab:exam_ic418} presents a demonstrative output of PyEMILI's identification of an emission line observed at 3918.93\,{\AA} (wavelength corrected for the $+$68.6 km\,s$^{-1}$ radial velocity of the PN as determined from the \ion{H}{ii} Balmer and Paschen lines) detected in the echelle spectrum of IC\,418, showing the top 15 candidate IDs.  This emission line has been manually identified by \citet{2003sharpee} as \ion{C}{ii} M4 2s$^{2}$\,3p\,$^{2}$P$^{\rm o}_{1/2}$--2s$^{2}$\,4s\,$^{2}$S$_{1/2}$ $\lambda$3918.97, which is also ranked as ``A'' by PyEMILI.  The candidate ID \ion{C}{ii} $\lambda$3918.97 has the highest predicted template flux value among all candidates as calculated using the available effective recombination coefficient and receives a score of 0 using the predicted template flux criteria (component $F$=0).  Additionally, the other two $IDI$ components of this \ion{C}{ii} transition possess the greatest conditions based on the wavelength difference (within 1$\sigma$) and the potential existence of the other fine-structure component of this transition, \ion{C}{ii} M4 2s$^{2}$\,3p\,$^{2}$P$^{\rm o}_{3/2}$--2s$^{2}$\,4s\,$^{2}$S$_{1/2}$ $\lambda$3920.69, with a similar velocity difference of $-$3.5\,km\,s$^{-1}$.  Consequently, the $IDI$ of \ion{C}{ii} $\lambda$3918.97 is determined to be 1, given that the three component scores are $W$=0, $F$=0, and $M$=1.  The second highest-ranking ID of this observed line at 3918.93\,{\AA} is \ion{N}{ii} M17 2s$^{2}$\,2p\,3p\,$^{1}$P$_{1}$--2s$^{2}$\,2p\,3d\,$^{1}$P$^{\rm o}_{1}$ $\lambda$3919.00, with $IDI$=3 due to a ``0/0'' multiplet-check condition and a predicted template flux within 1/100 relative to the largest predicted template flux; the component scores of this candidate ID are $W$=0, $F$=1, and $M$=2.  In such conditions, preference is given to the ID of \ion{C}{ii} $\lambda$3918.97 as the correct identification; it is less likely to be \ion{N}{ii} $\lambda$3919.00, given its predicted template flux is only about one percent of that of \ion{C}{ii} $\lambda$3918.97. 

To perform a more objective comparison between EMILI and PyEMILI, we use the results of manual identifications from \citet[][Table\,3 therein]{2003sharpee} as a benchmark reference.  Through careful scrutinization and interrogation of huge volumes of spectroscopic databases, we found a few mis-identifications, particularly for the faint emission lines, in the line table of \citet{2003sharpee}; this is inevitable due to incompleteness of the atomic transition database used by EMILI.  However, these mis-identifications have little impact on the general results, given the huge number of faint emission lines in this PN.  For each emission line in IC\,418, we checked its manual identification given by \citet{2003sharpee} against the identifications, as well as the rankings (A, B, C, or D) of the IDs, assigned by PyEMILI and EMILI, and count the numbers of rankings.  In cases where an observed line has multiple potential identifications, we select the highest ranking of these potential IDs for counting purposes.  For more than 60\% of the emission lines in the input line list of IC\,418, PyEMILI and EMILI gives the same ID as the manual identification of \citet{2003sharpee} and both assigned the ID with ranking A.  The identification of the observed line at 3918.93\,{\AA} in Table\,\ref{tab:exam_ic418} is one such example.  The detailed output of PyEMILI in the original format is accessible via the project website (see the {\sc url} in Section\,2.1). 

Table\,\ref{tab:ic418} provides number statistics for different rankings of the IDs assigned by EMILI and PyEMILI, compared to the results of manual identifications of \citet{2003sharpee} for IC\,418.  Out of the 806 lines detected in the spectrum of IC\,418 \citep{2004sharpee}, 696 were manually identified by \citet{2003sharpee} and 110 were unidentified.  PyEMILI has significantly increased the number of emission lines with the A-ranking IDs from 75\%, as given by EMILI, to now $\sim$91\%, and reduced the number of the lines that lack rankings in comparison to EMILI.  We re-identified 28 emission lines in IC\,418 to which we thought PyEMILI assigned more reliable IDs than the manual identifications in \citet{2003sharpee}.  Among the 110 unidentified emission lines in IC\,418, PyEMILI identified 38 which are deemed to be rigorous identifications.  The emission-line identifications of PyEMILI that are different from the manual identifications of \citet{2003sharpee} are listed in Table\,\ref{tab:addic418}, where the emission lines newly identified by PyEMILI (and unidentified in \citealt{2003sharpee}) are also presented.  The majority of the lines in Table\,\ref{tab:addic418} receive ranking ``A'' from PyEMILI.

\begin{longrotatetable}
\begin{deluxetable*}{cccccccccccc}
\tablecaption{Amended and Supplemented Identifications of the Emission Lines in IC\,418 \label{tab:addic418}}
\tablehead{\colhead{$\lambda_{\mathrm{obs}}$} & \colhead{$I_\lambda/I_{\mathrm{H\beta}}$} & \colhead{$\mathrm{ID_1}$} & \colhead{$\lambda_{\mathrm{lab1}}$} & \colhead{$\mathrm{Trans_1}$}  & \colhead{$\mathrm{ID_2}$} & \colhead{$\lambda_{\mathrm{lab2}}$} & \colhead{$\mathrm{Trans_2}$}  & \colhead{$\mathrm{Electron\,Configurations}$} & \colhead{$g_\mathrm{l}$} & \colhead{$g_\mathrm{u}$} & \colhead{$\mathrm{Notes}$}\\ 
\colhead{(\AA)} & \colhead{} & \colhead{} & \colhead{(\AA)} &  \colhead{(Lower--Upper)} & \colhead{} & \colhead{(\AA)} & \colhead{(Lower--Upper)} & \colhead{(Lower--Upper)} & \colhead{} & \colhead{} & \colhead{}\\
\colhead{(1)} & \colhead{(2)} & \colhead{(3)} & \colhead{(4)} & \colhead{(5)} & \colhead{(6)} & \colhead{(7)} & \colhead{(8)} & \colhead{(9)} & \colhead{(10)} & \colhead{(11)} & \colhead{(12)} }
\startdata
3574.78&4.68E-05&\ion{O}{ii}&3574.85&$\mathrm{^2D-{^2}P^o}$&\ion{Ne}{ii}&3574.61&$\mathrm{^2D-{^2}F^o}$&$\mathrm{2s^2.2p^4.(^1D).3s-2s^2.2p^4.(^1D).3p}$&4&6&ERC\\
3601.33&2.73E-05&…&…&…&\ion{Ne}{ii}&3601.06&$\mathrm{^4D-{^4}P^o}$&$\mathrm{2s^2.2p^4.(^3P).3d-2s^2.2p^4.(^3P).5p}$&8&6&ERC\\
3801.36&6.97E-05&…&…&…&\ion{S}{ii}&3801.39&$\mathrm{^2P^o-{^2}D}$&$\mathrm{3s^2.3p^2.(^1D).4p-3s^2.3p^2.(^3P).5d}$&4&4&TP\\
3923.20&4.34E-05&…&…&…&\ion{S}{ii}&3923.45&$\mathrm{^2D^o-{^2}F}$&$\mathrm{3s^2.3p^2.(^3P).4p-3s^2.3p^2.(^3P).4d}$&4&6&TP\\
3924.01&3.55E-05&…&…&…&\ion{S}{ii}&3924.06&$\mathrm{^2D^o-{^2}P}$&$\mathrm{3s^2.3p^2.(^1D).4p-3s^2.3p^2.(^1D).4d}$&6&4&TP\\
4015.93&4.09E-05&…&…&…&\ion{C}{ii}&4015.72&$\mathrm{^2D-{^2}F^o}$&$\mathrm{2s^2.4d-2s^2.11f}$&6&6&TP\\
4065.23&5.48E-05&\ion{Fe}{iii}&4065.25&$\mathrm{^5H^o-{^5}G}$&\ion{O}{ii}]&4065.05&$\mathrm{^4F-F[3]^o}$&$\mathrm{2s^2.2p^2.(^3P).3d-2s^2.2p^2.(^3P).4f.F^o}$&10&8&Mult, ERC\\
4084.67&4.73E-05&\ion{Fe}{i}&4084.49&$\mathrm{z^5F^o-g^5D}$&\ion{O}{ii}]&4084.65&$\mathrm{^4P^o-{^2}F}$&$\mathrm{2s^2.2p^2.(^3P).3p-2s^2.2p^2.(^3P).3d}$&6&8&Mult, ERC\\
4096.51&2.83E-05&[\ion{Fe}{iii}]&4096.68&$\mathrm{^5D-{^3}G}$&\ion{O}{ii}]&4096.53&$\mathrm{^4P^o-{^2}F}$&$\mathrm{2s^2.2p^2.(^3P).3p-2s^2.2p^2.(^3P).3d}$&4&6&Blend, Mult, ERC\\
4108.45&5.48E-05&…&…&…&\ion{O}{ii}&4108.75&$\mathrm{^4F-G[3]^o}$&$\mathrm{2s^2.2p^2.(^3P).3d-2s^2.2p^2.(^3P).4f.G^o}$&8&8&ERC\\
4161.14&2.16E-05&…&…&…&\ion{N}{ii}&4161.15&$\mathrm{^3D^o-D[3/2]}$&$\mathrm{2s^2.2p.(^2P^o).3d-2s^2.2p.(^2P^o).4f.D}$&5&3&ERC\\
4280.55&2.55E-05&\ion{Fe}{ii}]&4280.54&$\mathrm{z^2P^o-{^4}P}$&\ion{Ne}{ii}&4280.58&$\mathrm{^4F-1[4]^o}$&$\mathrm{2s^2.2p^4.(^3P).3d-2s^2.2p^4.(^3P\left \langle1\right \rangle).4f}$&10&10&TP\\
4287.55&8.29E-05&[\ion{Co}{ii}]&4287.50&$\mathrm{a^3F-b^3P}$&[\ion{Fe}{ii}]&4287.39&$\mathrm{a^6D-a^6S}$&$\mathrm{3d^6.(^5D).4s-3d^5.4s^2}$&10&6&Mult\\
4390.55&5.89E-05&…&…&…&\ion{Mg}{ii}&4390.57&$\mathrm{^2P^o-{^2}D}$&$\mathrm{2p^6.4p-2p^6.5d}$&4&6&TP\\
4562.64&4.14E-04&…&…&…&[\ion{Mg}{i}]&4562.60&$\mathrm{^1S-{^3}P^o}$&$\mathrm{3s^2-3s.3p}$&1&5&Mult\\
4654.48&3.14E-05&\ion{O}{i}&4654.56&$\mathrm{^5P-{^5}D^o}$&\ion{N}{ii}]&4654.53&$\mathrm{^1P^o-{^3}P}$&$\mathrm{2s^2.2p.(^2P^o).3s-2s^2.2p.(^2P^o).3p}$&3&5&Mult, ERC\\
4667.25&2.80E-05&[\ion{Fe}{iii}]&4667.11&$\mathrm{^5D-{^3}F^4}$&\ion{N}{ii}]&4667.21&$\mathrm{^1P^o-{^3}P}$&$\mathrm{2s^2.2p.(^2P^o).3s-2s^2.2p.(^2P^o).3p}$&3&3&Blend, Mult, ERC\\
4674.90&2.12E-05&…&…&…&\ion{N}{ii}]&4674.91&$\mathrm{^1P^o-{^3}P}$&$\mathrm{2s^2.2p.(^2P^o).3s-2s^2.2p.(^2P^o).3p}$&3&1&Mult, ERC\\
4710.01&6.69E-05&…&…&…&\ion{O}{ii}]&4710.01&$\mathrm{^2D^o-{^4}D}$&$\mathrm{2s^2.2p^2.(^3P).3p-2s^2.2p^2.(^3P).3d}$&4&6&ERC\\
4716.33&3.72E-05&[\ion{Fe}{iii}]&4716.43&$\mathrm{^3F^4-{^1}F}$&\ion{S}{ii}&4716.27&$\mathrm{^4P-{^4}S^o}$&$\mathrm{3s^2.3p^2.(^3P).4s-3s^2.3p^2.(^3P).4p}$&4&4&Mult\\
4752.89&1.72E-04&…&…&…&\ion{O}{ii}]&4752.69&$\mathrm{^2D^o-{^4}D}$&$\mathrm{2s^2.2p^2.(^3P).3p-2s^2.2p^2.(^3P).3d}$&6&6&ERC\\
4789.59&2.72E-04&…&…&…&[\ion{F}{ii}]&4789.45&$\mathrm{^3P-{^1}D}$&$\mathrm{2s^2.2p^4-2s^2.2p^4}$&5&5&TP\\
5172.79&4.89E-05&\ion{N}{ii}&5172.97&$\mathrm{^5D^o-{^5}F}$&\ion{Mg}{i}&5172.68&$\mathrm{^3P^o-{^3}S}$&$\mathrm{3s.3p-3s.4s}$&3&3&Blend, Mult\\
5183.81&4.94E-05&…&…&…&\ion{Mg}{i}&5183.60&$\mathrm{^3P^o-{^3}S}$&$\mathrm{3s.3p-3s.4s}$&5&3&Mult\\
5345.94&3.50E-05&…&…&…&[\ion{Kr}{iv}]&5346.02&$\mathrm{^4S^o-{^2}D^o}$&$\mathrm{4s^2.4p^3-4s^2.4p^3}$&4&6&Mult\\
5400.55&2.85E-05&…&…&…&\ion{Ne}{i}&5400.56&$\mathrm{^{3/2}[3/2]^o-{^{1/2}}[1/2]}$&$\mathrm{2s^2.2p^5.(^2P^o\left \langle3/2\right \rangle).3s-2s^2.2p^5.(^2P^o\left \langle1/2\right \rangle).3p}$&3&1&TP\\
5662.34&2.52E-05&…&…&…&\ion{C}{ii}&5662.46&$\mathrm{^4P^o-{^4}S}$&$\mathrm{2s.2p.(^3P^o).3s-2s.2p.(^3P^o).3p}$&6&4&Mult\\
5747.29&3.28E-05&\ion{O}{ii}&5747.33&$\mathrm{^2D^o-{^2}F}$&\ion{N}{ii}]&5747.30&$\mathrm{^1P^o-{^3}D}$&$\mathrm{2s^2.2p.(^2P^o).3s-2s^2.2p.(^2P^o).3p}$&3&5&Mult, ERC\\
5767.46&1.25E-05&\ion{S}{ii}&5767.42&$\mathrm{^4D^o-{^4}P}$&\ion{N}{ii}]&5767.45&$\mathrm{^1P^o-{^3}D}$&$\mathrm{2s^2.2p.(^2P^o).3s-2s^2.2p.(^2P^o).3p}$&3&3&Mult, ERC\\
5832.96&2.84E-05&…&…&…&\ion{C}{iii}&5833.20&$\mathrm{^1F^o-{^1}G}$&$\mathrm{2s.5f-2s.7g}$&7&9&TP\\
5867.73&3.50E-05&\ion{Al}{ii}&5867.89&$\mathrm{^3D-{^3}F^o}$&\ion{Si}{ii}&5867.43&$\mathrm{^4P^o-{^4}P}$&$\mathrm{3s.3p.(^3P^o).4s-3s.3p.(^3P^o).4p}$&4&2&Mult\\
&&&&&[\ion{Kr}{iv}]&5867.74&$\mathrm{^4S^o-{^2}D^o}$&$\mathrm{4s^2.4p^3-4s^2.4p^3}$&4&4&Mult\\
5915.61&3.37E-05&…&…&…&\ion{Si}{ii}&5915.19&$\mathrm{^4P^o-{^4}P}$&$\mathrm{3s.3p.(^3P^o).4s-3s.3p.(^3P^o).4p}$&6&4&Mult\\
5991.09&9.44E-06&\ion{C}{i}&5990.98&$\mathrm{^3D-{^3}D^o}$&\ion{O}{i}&5991.31&$\mathrm{^3D-{^3}P^o}$&$\mathrm{2s^2.2p^3.(^2D^o).3p-2s^2.2p^3.(^2D^o).4d}$&3&1&TP\\
6379.65&8.63E-06&\ion{O}{ii}&6379.58&$\mathrm{D[3]^o-{^2}[2]}$&\ion{N}{ii}]&6379.62&$\mathrm{^3P^o-{^1}P}$&$\mathrm{2s^2.2p.(^2P^o).3s-2s^2.2p.(^2P^o).3p}$&3&3&ERC\\
6725.25&1.54E-05&…&…&…&\ion{He}{i}&6725.25&$\mathrm{^3S-{^3}P^o}$&$\mathrm{1s.3s-1s.23p}$&3&5&ERC, Abun\\
6826.87&3.30E-04&…&…&…&[\ion{Kr}{iii}]&6826.70&$\mathrm{^3P-{^1}D}$&$\mathrm{4s^2.4p^4-4s^2.4p^4}$&5&5&TP\\
6829.90&1.04E-05&[\ion{Fe}{ii}]&6830.04&$\mathrm{a^4D-a^6S}$&\ion{Si}{ii}&6829.83&$\mathrm{^2P^o-{^2}D}$&$\mathrm{3s^2.5p-3s^2.6d}$&4&6&TP\\
7902.15&1.33E-05&…&…&…&\ion{He}{i}&7902.26&$\mathrm{^3P^o-{^3}D}$&$\mathrm{1s.3p-1s.34d}$&5&7&Abun\\
7906.00&3.38E-05&…&…&…&\ion{He}{i}&7905.91&$\mathrm{^3P^o-{^3}D}$&$\mathrm{1s.3p-1s.33d}$&5&7&Abun\\
7932.98&1.73E-05&…&…&…&\ion{He}{i}&7932.41&$\mathrm{^3P^o-{^3}S}$&$\mathrm{1s.3p-1s.28s}$&5&3&Abun\\
8064.89&7.15E-06&…&…&…&\ion{He}{i}&8064.96&$\mathrm{^3P^o-{^3}S}$&$\mathrm{1s.3p-1s.18s}$&5&3&ERC, Abun\\
8213.89&2.59E-05&[\ion{Fe}{iii}]&8214.24&$\mathrm{^3D-{^1}F}$&\ion{Mg}{ii}&8213.99&$\mathrm{^2P^o-{^2}S}$&$\mathrm{2p^6.4p-2p^6.5s}$&2&2&Mult\\
8233.20&1.17E-04&\ion{Mg}{ii}&8233.19&$\mathrm{^2G-{^2}H^o}$&\ion{H}{i}&8233.21&$\mathrm{3-50}$&$\mathrm{3*-50*}$&*&*&Abun\\
&&&&&\ion{O}{i}&8233.00&$\mathrm{^3D^o-{^3}D}$&$\mathrm{2s^2.2p^3.(^2D^o).3s-2s^2.2p^3.(^2D^o).3p}$&3&3&TP\\
8234.50&2.42E-04&\ion{Mg}{ii}&8234.64&$\mathrm{^2P^o-{^2}S}$&\ion{H}{i}&8234.43&$\mathrm{3-49}$&$\mathrm{3*-49*}$&*&*&Blend, Abun\\
8235.77&1.21E-04&…&…&…&\ion{H}{i}&8235.74&$\mathrm{3-48}$&$\mathrm{3*-48*}$&*&*&Abun\\
8237.12&1.47E-04&…&…&…&\ion{H}{i}&8237.13&$\mathrm{3-47}$&$\mathrm{3*-47*}$&*&*&Abun\\
8238.62&1.23E-04&\ion{S}{ii}&8238.59&$\mathrm{^2D-{^2}P^o}$&\ion{H}{i}&8238.61&$\mathrm{3-46}$&$\mathrm{3*-46*}$&*&*&Abun\\
8240.20&2.25E-04&\ion{Fe}{iii}&8240.72&$\mathrm{^5D-{^5}D^o}$&\ion{H}{i}&8240.19&$\mathrm{3-45}$&$\mathrm{3*-45*}$&*&*&Abun\\
8242.03&2.68E-04&…&…&…&\ion{H}{i}&8241.88&$\mathrm{3-44}$&$\mathrm{3*-44*}$&*&*&Abun\\
8264.34&9.36E-04&\ion{He}{i}&8264.56&$\mathrm{^3P^o-{^3}D}$&\ion{H}{i}&8264.28&$\mathrm{3-35}$&$\mathrm{3*-35*}$&*&*&Blend, Abun\\
8285.69&1.01E-04&\ion{He}{i}&8285.44&$\mathrm{^3P^o-{^3}S}$&\ion{C}{ii}&8285.70&$\mathrm{^2P^o-{^2}D}$&$\mathrm{2s^2.5p-2s^2.7d}$&2&4&Blend, Mult\\
8320.49&2.04E-05&\ion{Fe}{i}&8320.17&$\mathrm{c^3F-y^3G^o}$&\ion{He}{i}&8320.50&$\mathrm{^1D-{^1}F^o}$&$\mathrm{1s.3d-1s.24f}$&5&7&ERC, Abun\\
8339.34&5.21E-05&\ion{Ca}{i}&8339.21&$\mathrm{^1D-{^1}P^o}$&\ion{He}{i}&8340.00&$\mathrm{^1P^o-{^1}D}$&$\mathrm{1s.3p-1s.31d}$&4&2&Abun\\
8352.47&9.84E-06&…&…&…&\ion{He}{i}&8352.22&$\mathrm{^1P^o-{^1}S}$&$\mathrm{1s.3p-1s.29s}$&3&1&Abun\\
8653.42&1.71E-05&\ion{He}{i}&8653.39&$\mathrm{^3D^o-{^3}P}$&\ion{C}{iii}&8653.53&$\mathrm{^3D^o-{^3}P}$&$\mathrm{2s.5d-2s.6p}$&3&1&Blend, TP\\
8873.45&3.23E-05&\ion{C}{i}&8873.36&$\mathrm{^1D-{^1}P^o}$&\ion{C}{ii}&8873.4&$\mathrm{^2D-{^2}F^o}$&$\mathrm{2s^2.5d-2s^2.8f}$&6&6&ERC\\
8886.54&7.18E-06&…&…&…&\ion{O}{ii}&8886.31&$\mathrm{^4F-D[3]^o}$&$\mathrm{2s^2.2p^2.(^3P).4d-2s^2.2p^2.(^3P).5f.D^o}$&6&8&ERC\\
9052.45&1.70E-05&…&…&…&[\ion{Fe}{ii}]&9051.95&$\mathrm{a^4F-a^4P}$&$\mathrm{3d^7-3d^7}$&8&6&Mult\\
9143.13&2.24E-05&…&…&…&\ion{O}{ii}&9143.50&$\mathrm{^2D-F[3]^o}$&$\mathrm{2s^2.2p^2.(^1D).4d-2s^2.2p^2.(^1D).5f.F^o}$&6&8&TP\\
9223.64&6.37E-05&…&…&…&\ion{C}{ii}&9224.00&$\mathrm{^2F^o-{^2}G}$&$\mathrm{2s^2.5f-2s^2.8g}$&6&8&TP\\
9236.26&1.36E-04&…&…&…&\ion{C}{ii}&9236.82&$\mathrm{^2P^o-{^2}S}$&$\mathrm{2s^2.4p-2s^2.5s}$&4&2&TP\\
9318.09&1.14E-04&…&…&…&\ion{C}{ii}&9317.20&$\mathrm{^2G-{^2}H^o}$&$\mathrm{2s^2.5g-2s^2.8h}$&8&10&TP\\
9805.38&5.22E-05&…&…&…&\ion{O}{ii}&9806.16&$\mathrm{D[2]^o-{^2}[3]}$&$\mathrm{2s^2.2p^2.(^3P).4f.D^o-2s^2.2p^2.(^3P\left \langle2\right \rangle).5g}$&4&6&ERC\\
\enddata
\tablecomments{The emission lines presented in this table have the PyEMILI identifications different from those by EMILI, and include the newly identified lines that were unidentified in \citet{2003sharpee}. \\
\smallskip\\
Columns (1) and (2): The observed wavelength (corrected for the radial velocity of IC\,418) and extinction-corrected line intensity adopted from \citet[][Table\,3 therein]{2003sharpee}. \\
Columns (3), (4) and (5): The emitting ion, the laboratory wavelength, and the transition, respectively, assigned by \citet[][see also Table\,3 therein]{2003sharpee}.  ``...'' indicates that the line was unidentified in \citet{2003sharpee}. \\
Columns (6), (7) and (8): Same as the meanings of Columns (3), (4) and (5), but assigned by PyEMILI. \\
Columns (9): Electron configurations of the lower and upper states of the transition ID assigned by PyEMILI. \\
Columns (10) and (11): The statistical weights of lower and upper levels, respectively, of Column (8). \\
Column (12): The reasons why the new ID (Columns 6--11) assigned by PyEMILI is more robust: 
ERC$-$ The predicted intensity of the new ID is calculated using the effective recombination coefficients collected for our atomic transition database (see Section\,\ref{ERC}), and is more consistent with the observed intensity.  Mult$-$ The new ID has other multiplet members present in the line list of IC\,418, identified either by us or by \citet{2003sharpee}.  TP$-$ For the permitted transitions, this implies higher transition probabilities; for the forbidden transitions, this means the transition probability is higher and the upper-level energy is lower (and thus easier to be collisionally excited).  Both cases lead to higher predicted intensities.  Abun$-$ A higher ionic abundance for the new ID.  Blend$-$ The primitive ID is also possible but has possible blended transition(s). \\
\smallskip\\
(This table is online available in the machine-readable format.)
}
\end{deluxetable*}
\end{longrotatetable}

\subsection{Hf\,2-2} 
\label{hf2_2}

\subsubsection{The Abundance Discrepancy Problem}
\label{hf2_2:sec1}

Galactic PN Hf\,2-2 is renowned for numerous prominent ORLs from heavy-element ions (mainly \ion{O}{ii}, \ion{N}{ii} and \ion{Ne}{ii}) detected in the optical spectrum, as well as very high abundance discrepancy factor (ADF$\sim$70; \citealt{2006MNRAS.368.1959L}).  The abundance discrepancy factor was defined as the ratio of the ionic/elemental abundance derived from the ORLs to that (of the same ion/element) derived from the CELs.  The ``abundance discrepancy'' is a long-standing dichotomy whereby the ionic (and also elemental) abundances of C, N, O and Ne relative to hydrogen determined from ORLs are systematically higher than those derived for the same ions from the much brighter CELs.  The abundance discrepancy was first discovered through optical spectroscopy of Galactic \ion{H}{ii} regions and later in many PNe \citep[e.g.][]{Peimbert1967,Rubin1989,2000liu}.  Today the abundance discrepancy has been found and ADF values derived in 150 Galactic PNe \citep{Wesson2018}. 

A number of postulations had been raised to explain this discrepancy \citep[see the review of][]{2010arXiv1001.3715L}.  Provoked by high ADFs observed in PNe, \citet{2000liu} suggested a bi-abundance nebular model for PNe, where a previously unknown component of cold, metal-rich ionized gas, possibly in the form of ``pockets'', exists in the ambient hot ($\sim$10$^{4}$~K), H-rich nebula; the ORLs of heavy-element ions are emitted predominantly from these cold, metal-rich regions.  The bi-abundance nebular model, well supported by recent spectroscopic observations of PNe \citep[e.g.][]{2022MNRAS.510.5444G,2022AJ....164..243R}, naturally explains the large ADF values observed in many Galactic PNe \citep{2010arXiv1001.3715L}, and if true, the astrophysical origin(s) of this cold, metal-rich component remains unknown.  Recent discovery of a link between the PN central star binarity and extreme ADF values in PNe might help to shed light on this classical problem \citep{Wesson2018}.

\subsubsection{The Typicality of Hf\,2-2} 
\label{hf2_2:sec2}

Hf\,2-2 is an archetypal PN to investigate the abundance discrepancy problem given its rich optical recombination spectrum, even with the medium spectral resolution \citep[][]{2006MNRAS.368.1959L}.  The extreme strengths of the recombination lines of heavy-element ions  in its optical spectrum and the very sharp Balmer discontinuity of nebular continuum at 3646\,{AA} (a.k.a.\ the Balmer jump), which yields a very low electron temperature of $\sim$900\,K \citep[][see Figure\,1 therein]{2006MNRAS.368.1959L}, seem to well support the presence of a cold, metal-rich component of gas in Hf\,2-2, although at this point it is impossible to favour any scenario for the abundance discrepancy due to the lack of sufficient observational constraints.  These characteristics in nebular spectroscopy made Hf\,2-2 of particular interest \citep[e.g.][]{2010arXiv1001.3715L,Storey2014}.  Contrary to the detailed spectroscopy of the nebula, the central star of Hf\,2-2 is not yet thoroughly studied given the limited amount of existing data, although it has been reported as a periodic photometric variable with a period of 0.398571~d\citep{Lutz1998}, and preliminary ranges of the binary system parameters (effective temperature of the companion, radii of the central star and its companion, mass ratio, orbital radius, system inclination, etc.) were derived in a binary assumption \citep{Hillwig2016}.

\subsubsection{The VLT/UVES Spectroscopy of Hf\,2-2} 
\label{hf2_2:sec3}

Deep echelle spectroscopy of Hf\,2-2 was obtained in 2002, along with another two high-ADF Galactic PNe NGC\,6153 and M\,1-42, using the UVES spectrograph on the ESO 8.2\,m VLT (prop.~ID: 69.D-0174(A), PI: J.\ Danziger).  Four cross-dispersers (CD\#1, CD\#2, CD\#3 and CD\#4) from two dichroics of UVES\footnote{\url{http:///www.eso.org/sci/facilities/paranal/instruments/uves/index.html}}, covering a wavelength range of $\sim$3040--11,000\,{\AA}, and a slit width of 2\arcsec\ were used in the observations.  This instrument setup secures a spectral resolution of $R\sim$20,000.  These echelle spectra were originally reduced using the UVES pipeline, and the sky background was removed from the spectra using ESO {\sc midas}\footnote{{\sc midas} is developed and distributed by the European Southern Observatory.} \citep{2016MNRAS.461.2818M}.  More details of the observations, as well as spectral analysis, are described in \citet{2016MNRAS.461.2818M}.  In total, 310 emission lines were measured in the spectrum of Hf\,2-2, and identified using EMILI by \citet[][Section\,4 therein]{2016MNRAS.461.2818M}; these identifications are compared with our PyEMILI output for this PN in this section.

\subsubsection{Measurements and Identification of Emission Lines in Hf\,2-2} 
\label{hf2_2:sec4}

Before we proceed to present the line identifications made by PyEMILI for Hf\,2-2 and compare them with those of EMILI, it should be mentioned that the UVES pipeline did not provide absolute flux calibration (e.g.\ only the relative flux calibration was made in data reduction).  Moreover, the line fluxes of Hf\,2-2 reported in \citet{2016MNRAS.461.2818M}, who utilized the UVES pipeline-reduced spectra for emission line measurements, have been found to be unreliable probably due to problems with flux calibration, extinction correction and/or telluric absorptions \citep{2022MNRAS.510.5444G}.  In order to obtain accurate and absolute flux calibration of the UVES spectra of the three high-ADF PNe NGC\,6153, Hf\,2-2 and M\,1-42, we had carried out data reduction over again, systematically and carefully, from the very beginning, using the UVES spectra of the spectrophotometric standard stars retrieved from the ESO archive.  These newly reduced, fully calibrated UVES spectra and careful measurements of all the emission lines in the three high-ADF PNe will be reported separately (Huang, Fang \& Liu, in preparation). 

Therefore, we undertook a thorough effort to check the emission features in the UVES 2D frames, and manually re-measure the emission lines in the extracted 1D spectra.  Unfortunately, the echelle slit used in conjunction with the four cross-dispersers of UVES are not identical in length, and do not cover exactly the same region of the PN.  We inspect each order of the UVES 2D spectra by eyes and trimmed the data to retain only the same nebular regions along the slit (excluding the central star).  This is to ensure that all the emission lines we measured come from the same nebular region of Hf\,2-2.  For cases where an emission line is seriously blended with skyline emission, it is difficult to remove the skylines correctly without a sky-emission spectrum obtained with the same telescope and at the same observing time.  When checking through the UVES 2D spectra for the possible cases of nebular line-sky line blending, we use the flux-calibrated, high-resolution atlas of night-sky optical-NIR emission from UVES give by \citet{2003skylines} as a reference.  The emission lines that suffer the skyline-blending issue are not included in our line list of Hf\,2-2.  Fortunately, we found that in the UVES spectrum of Hf\,2-2, only $\sim$20 nebular emission lines are blended with skylines, and they are all faint and mostly in the red wavelength region at $>$7000\,{\AA}. 

The 1D spectra were then extracted by combining the rows of CCD pixels occupied by nebular emission in the UVES 2D spectra, excluding the central star.  Most emission lines in the 1D spectra of Hf\,2-2 are single-peaked, and were fitted with Gaussian profiles to obtain the central wavelengths (i.e.\ the observed wavelengths of emission lines).  For an emission line with the double-peak profile (due to nebular expansion), two Gaussian profiles were used to fit the two emission peaks; then an average wavelength of the two emission peaks is adopted as the observed wavelength of this line.  Line fluxes were then measured via integration over the line profile.  Some strong emission lines in the overlapping wavelength region covered by two cross-dispersers show slight differences in the flux value.  In order to reduce such difference in line flux and bring the adjacent orders of UVES spectra to the same level, we use the scale factors that were derived based on measurements of the emission lines in the overlapping regions covered by the cross-dispersers CD\#1--\#2 and CD\#2--\#3.

\begin{deluxetable}{cccc}[t]
\tablecaption{Statistics of the Emission Lines in Hf\,2-2 Identified by PyEMILI \label{tab:Fluxerr}}
\setlength{\tabcolsep}{0.25cm}
\tablehead{\colhead{Intensity Range} & \colhead{Number} & \colhead{Int. Error} & \colhead{Fraction of A-ranking}\\
\colhead{$I_\lambda/I_{\mathrm{H\beta}}$} & \colhead{} & \colhead{(\%)} & \colhead{(\%)} } 
\startdata
$\leq$0.001 &  102 &  6.53 &  96.1\\
0.001--0.01 &  246 &  4.36 &  98.0\\
  0.01--0.1 &   56 &  1.07 & 100  \\
  $\geq$0.1 &   14 &  0.08 & 100  \\
\enddata
\tablecomments{Number: the total number of observed lines in the corresponding line intensity range. Intensity error: the average value of line intensity error. A ranking rate: The percentage of lines with ``A'' ranking.}
\end{deluxetable}

Emission-line fluxes measured in the wavelength region $\sim$3730--5000\,{\AA} covered by CD\#3 were multiplied by a factor of 1.24.  The cross-dispersers CD\#3 and CD\#4 do not overlap in wavelength, and we assumed that the relative flux calibration of the UVES spectra is generally acceptable.  For an emission line in the overlapping wavelength region covered by two cross-dispersers, its flux is an average of the two values separately measured.  The flux error of an emission line was estimated using the standard deviation of nearby local continuum multiplied by a square root of the pixel number covered by this line.  Table\,\ref{tab:Fluxerr} presents the average errors in line flux and number statistics in the different ranges of relative intensity ($I_\lambda$/$I_{{\rm H}\beta}$) of emission lines.  In the range $I_\lambda$/$I_{{\rm H}\beta}$ $\leq$0.001, there are three emission lines with flux errors $>$20\%, possibly affected by low S/N ratios; the measurement errors of the three lines are not included in the calculation of the average flux error.  The extinction parameter $c$(H$\beta$) to Hf\,2-2 was derived using the \ion{H}{i} lines of the Balmer series up to the principal quantum number $n$=9 of the upper level.  Based on the measured line fluxes, we derived an average value of $c$(H$\beta$) = 0.38, using the effective recombination coefficients of the \ion{H}{i} lines calculated by \citet[][at $T_{\rm e}$ = 10,000\,K and $N_{\rm e}$ = 10,000\,cm$^{-3}$]{HIcoe} in the Case~B assumption.  All the line fluxes were then extinction-corrected using the $c$(H$\beta$). 

In total, we detected and measured 418 emission lines in the long-exposure (2$\times$1800\,s) UVES spectrum of Hf\,2-2; in contrast, only 310 emission lines were detected by \citet{2016MNRAS.461.2818M} in the same spectrum.  Of the 418 emission lines we detected, 411 were manually identified and checked against previous spectroscopic observations \citep{2006MNRAS.368.1959L,2016MNRAS.461.2818M,2022MNRAS.510.5444G}; the remaining seven lines were considered as dubious features due to faintness as well as poor S/N's.  We first carried out manual identifications of emission lines by checking each line against the atomic transition database, using the published line tables of Hf\,2-2 as references.  We then run PyEMILI, using the measured line list of Hf\,2-2 as the input.  After that, we checked the PyEMILI output for the \emph{IDI} and ranking of the ID assigned by manual identification for each emission line.  A statistics of the rankings is presented in Table\,\ref{tab:hf2-2_J0608}. 

The observed wavelengths ($\lambda_{\rm obs}$) of emission lines were corrected for the average radial velocity $v_{\rm ral}$ = 44.90$\pm$2.33\,km\,s$^{-1}$ as calculated using the measurements of the \ion{H}{i} Balmer and Paschen lines.  The 1$\sigma$ wavelength uncertainties of the emission lines were all set to be 10\,km\,s$^{-1}$, thanks to very high resolution of the UVES spectra ($R\sim$20,000).  Comparing our measurements of emission lines in Hf\,2-2 with those of \citet{2022MNRAS.510.5444G}, we found that nearly all the dereddened intensities of the \ion{H}{i} and \ion{He}{i} lines, relative to H$\beta$, we measured are slightly lower.  The H$\alpha$ flux we measured is two times lower than that given by \citet{2022MNRAS.510.5444G}; thus We did not include H$\alpha$ in the calculation of the extinction parameter $c$(H$\beta$).  The line fluxes of the low-ionization species such as [\ion{N}{ii}] $\lambda\lambda$6548,\,6583 and [\ion{S}{ii}] $\lambda\lambda$6716,\,6731, are also two to four times lower than those of \citet{2022MNRAS.510.5444G}.  However, the line fluxes of relatively high-ionization species such as \ion{He}{ii} $\lambda$4686, [\ion{Ar}{iv}] $\lambda\lambda$4711,\,4740 and [\ion{Cl}{iv}] $\lambda\lambda$7531,\,8046, are all higher than those from the literature. 

We attribute these discrepancies in line fluxes to the different nebular regions covered by the long/echelle slits in different spectroscopic observations.  The emission-line fluxes reported by \citet{2022MNRAS.510.5444G}, who carried out integral field unit (IFU) spectroscopy of Hf\,2-2 using the Multi-Unit Spectroscopic Explorer (MUSE) on the VLT, is integrated over the entire nebular region of Hf\,2-2, whereas our VLT UVES 1D spectra were extracted only $\lesssim$3.5\arcsec\ from Hf\,2-2's centre core, where most of the high-ionization ions are expected to concentrate.  The aforementioned differences between our measured line fluxes and those reported in the literature will not be further discussed, since in this paper we only concern about the performance of PyEMILI in identification of the emission line list of Hf\,2-2.  A critical reappraisal and analyses of the UVES spectra will be presented in a separate paper (Huang, Fang \& Liu, in preparation). 

A number statistics of the emission-line identifications for Hf\,2-2 are summarised in Table\,\ref{tab:hf2-2_J0608}.  More than 95\% of the emission lines that are ranked as ``A'' by PyEMILI are consistent with our manual identifications.  Most of the remaining lines with lower rankings may be explained by the absence of multiplet members or low transition probabilities.  Three emission lines without any ranking might be \ion{O}{ii} lines, whose transition data do not exist in our atomic transition database.  If the results of line identifications are compared based on the line intensities, the fraction of the A-ranking lines with intensity $>$0.01 of H$\beta$ is 100\%, and $\sim$96\% for the line intensity $<$0.001 of H$\beta$ (see Table\,\ref{tab:Fluxerr}).  To conclude, PyEMILI's identifications of the emission lines in the high-dispersion UVES spectrum of Hf\,2-2 are generally very satisfactory.

\begin{deluxetable}{ccccc}[t]
\tablecaption{Number Statistics of PyEMILI's Line-identification Results
\label{tab:hf2-2_J0608}}
\setlength{\tabcolsep}{0.42cm}
\tablehead{\colhead{Rank} & \multicolumn{2}{c}{\underline{~~~~~~~~Hf\,2-2~~~~~~~~}} & \multicolumn{2}{c}{\underline{~~~~~~~~J0608~~~~~~~~~}} \\ 
\colhead{} & \colhead{(Number)} & \colhead{(\%)} & \colhead{(Number)} & \colhead{(\%)} } 
\startdata
A    & 402  & 97.8  & 257  & 91.4\\
B    &   6  &  1.5  &  15  &  5.3\\
C    &   0  &  0    &   5  &  1.8\\
D    &   0  &  0    &   1  &  0.4\\
None &   3  &  0.7  &   3  &  1.1\\
\enddata
\tablecomments{A total number of 418 emission lines were detected in the UVES echelle spectrum of Hf\,2-2, of which 411 were identified; the remaining seven lines are dubious features and not included in number statistics. 
For J0608, 281 out of the 312 detected lines were manually identified. \\
References of previous manual line-identifications: Hf\,2-2 -- \citet{2016MNRAS.461.2818M}, \citet{2022MNRAS.510.5444G}, and \citet{2006MNRAS.368.1959L}.
J0608 -- \citet{Williams2021}.} 
\end{deluxetable}

\subsection{UVQS\,J060819.93-715737.4} 
\label{j0608} 

As an effort to validate the extend capability of PyEMILI in identifying more types of emission-line objects other than PNe and \ion{H}{ii} regions, we also run the code on the optical high-dispersion spectrum of UVQS\,J060819.93-715737.4 (hereafter J0608), a low-mass, late-type [WC11] star recently discovered in the Large Magellanic Cloud (LMC) by \citep{2020ApJ...888...54M}.

\subsubsection{The Late-type [WR] Stars} 
\label{j0608:sec1}

In the Milky Way, there is a population of H-poor white dwarfs that exhibit broad emission lines from He, C, N, and/or O in their spectra, and they all are the central stars of PNe.  The PNe with H-poor central white dwarfs are believed to have experienced a very late thermal pulse \citep[VLTP,][]{1983ApJ...264..605I} while the central star was descending the white dwarf cooling track \citep[e.g.][]{Guerrero_2018}; these PNe are called the ``born-again'' PNe, and only several of them are discovered in the Milky Way.  During the VLTP event, the remnant helium envelope among the stellar upper layers reaches the critical mass to ignite helium fusion (a final helium shell flash, as dubbed by \citealt{1983ApJ...264..605I}) into carbon and oxygen \citep[e.g.][]{Herwig2005,MillerBertolami2006}.  So far, about 10\% of the central stars of Galactic PNe are H-poor, low-mass white dwarfs, which are denoted as [WR]-type (see explanation of this denotation in Section\,2.6.1).  \emph{J0608 belongs to such category of [WR] stars but is the first confirmed extragalactic late-[WC] star}.  The population and occurrence of [WR]-type white dwarfs in the LMC (and SMC) are an interesting research topic given that both Magellanic Clouds have metal-poor environments compared to the Milky Way.

\subsubsection{The MagE Echelle Spectroscopy of J0608} 
\label{j0608:sec2}

J0608 was serendipitously discovered by \citet{2020ApJ...888...54M} in a survey for \ion{C}{ii} emission-line stars in the LMC \citep{Margon_2020}.  We use the late-type [WC11] Wolf-Rayet star J0608 as an example to demonstrate how PyEMILI would perform on other emission-line objects.  The spectrum we use for line identifications was obtained on 2019 May 3, using the Magellan Echellette (MagE) Spectrograph on the 6.5\,m Baade Magellan telescope \citep{2020ApJ...888...54M}.  This MagE spectroscopy provides a broad wavelength coverage of $\sim$3100--11,000\,{\AA} with a spectral resolution of $\sim$1\,{\AA} \citep[][Figure\,1 therein]{2020ApJ...888...54M}.  Numerous strong \ion{C}{ii} emission lines are seen in the spectrum, with $\lambda$4267, $\lambda$6580 and $\lambda$7236 being the most prominent.  The prevalence of \ion{C}{ii} emission lines in the spectrum of J0608 indicates its subclassification as very late-type [WC11] (probably the latest, lowest-excitation end of the [WC] sequence), which is well supported by the lack of both \ion{C}{iv} (e.g.\ the $\lambda\lambda$5801.33,\,5811.98 doublet) and \ion{He}{ii} emission \citep{1998MNRAS.296..367C}.  Moreover, many emission lines have prominent P~Cygni profiles (i.e.\ emission with blueward absorption), implying mass loss in the form of stellar winds. 

Although no obvious surrounding nebula was detected, the emission lines of \ion{O}{ii}, \ion{N}{ii} and \ion{S}{ii} in the spectrum of J0608 suggests the presence of a low-excitation nebula \citep{2020ApJ...888...54M}; thus J0608 is probably the [WC]-type central star of an evolved PN.  This is consistent with the current findings that all known Galactic late-type [WC] stars are the central stars of PNe.  The absence of detectable nebulosity around J0608 is probably due to the distance \citep[$\sim$50~kpc,][]{Pietrzyski2013} to the LMC, where most of the PNe therein have angular sizes $\lesssim$1\arcsec.

\subsubsection{Identification of Emission Lines in J0608} 
\label{j0608:sec3}

Careful \emph{manual} identification of the emission lines detected in the MagE echelle spectrum of J0608, utilizing atomic transition database, was carried out by \citep{Williams2021}, showing an exceptionally large number of \ion{C}{ii} emission lines originating from autoionizing levels.  Preliminary identification of emission lines and earlier analysis of the MagE echelle spectrum of J0608 were made by \citet{2020ApJ...888...54M}, followed by a more complete identification of the lines by \citet{Williams2021}. 
In this section, we use the latter identification of \citet{Williams2021} as a benchmark for our test run of the PyEMILI code on the spectrum of J0608. 

We re-measured all the spectral lines as listed in \citet[][Table\,1 therein]{Williams2021}, using the MagE echelle spectrum of J0608 provided by the author (R.\ Williams).  For the radial-velocity correction of the measured spectral lines, we adopted an average value of $v_{\rm ral}$ = 232.1\,km\,s$^{-1}$ for pure emission lines and 211.6\,km\,s$^{-1}$ for pure absorption lines as given by \citet{Williams2021}.  For many spectral lines with P-Cygni profiles, we assume that the emission and the absorption components of each line have a constant velocity difference of 20.5\,km\,s$^{-1}$ (= 232.1$-$211.6\,km\,s$^{-1}$), and we fit the two components empirically using two Gaussian profiles (i.e., of the emission and the absorption).  The Gaussian-fitted central wavelength of the emission component of a P-Cygni profile is adopted as the observed wavelength of this spectral line; it can be slightly blueward with respect to the observed emission peak of the P-Cygni profile because of the blueward absorption to the left wing of an emission feature, which is assumed to have a Gaussian profile. 

In our final line list, the observed wavelengths of all spectral lines were measured using the methods described above, and corrected for an average radial velocity based on measurements of multiple narrow emission lines in the MagE echelle spectrum of J0608.  As mentioned in section \ref{IDcriteria:3}, during the multiplet check, PyEMILI searches for possible detected fine-structure lines belonging to a multiplet only with the same line type, e.g.\ either the absorption lines or the emission lines, in the \textit{Input Line List}.  We did not correct the line fluxes for interstellar extinction since the measurement accuracy of the line fluxes is adequate for line identifications. 

In the calculation of the ionization structure parameter of the model used in PyEMILI, we typically utilize the table of PN-specific elemental abundances to derive the ionic abundances of various elements for the identification of emission lines in a PN (See Section~\ref{ISP}).  For other types of emission-line objects, the solar abundances \citep[e.g., of][]{2009ARA&A..47..481A} are used.  However, the significant discrepancies in some elemental abundances between the [WC11]-type stars and the solar values render both groups of abundances (the solar and the PNe) inappropriate.  Consequently, for the [WC] stars we have adopted the elemental abundances provided by \citet{2006WCabun}, specifically for He, C, O, N, Ne, Si, S, and Fe.  These revised abundances were subsequently incorporated into PyEMILI to facilitate ionic abundance calculations, which were then use to estimate the predicted template fluxes of the candidate IDs for an emission line being identified by PyEMILI.  For other elemental abundances, we adopt typical values of Galactic PNe. 

The MagE spectrum of J0608 indicates an extremely high carbon abundance (i.e.\ it is hydrogen-deficient).  The existence of several [\ion{O}{ii}], [\ion{N}{ii}], and [\ion{S}{ii}] emission lines in the spectrum suggests possible presence of an inapparent, low-excitation nebula surrounding the central star.  Thus some spectral lines contain radiation from both the central star and the nebula, making it somewhat complicated to determine the velocity structure of stellar winds.  Moreover, numerous absorption lines and P-Cygni profiles make analysis the velocity structure more complicated.  The MagE spectrum we re-measured has a spectral resolution of $\sim$1\,{\AA}.  Therefore, when using the criteria of wavelength agreement in the line identification for J0608, we reduced the weight of this criteria by designating a larger 1\,$\sigma$ wavelength uncertainty of 30\,km\,s$^{-1}$ for each observed line.  One disadvantage, if we adopted a larger wavelength uncertainty, is that more candidate IDs will be introduced, adding non-negligible complexity to the line identification. 

The manual identifications of the spectral lines in J0608 along with their corresponding $IDI$s and rankings assigned by PyEMILI are presented in Table\,\ref{j0608linelist}.  A summary of number statistics of the PyEMILI rankings of the line IDs given by the manual identification is presented in Table\,\ref{tab:hf2-2_J0608}.  A total of 312 spectral lines were detected in the MagE spectrum \citep{Williams2021}, of which 244 are emission lines and 68 are absorption lines; in total 281 lines were manually identified.  The majority of PyEMILI's identification results show a remarkable agreement with the manual identifications of \citet{Williams2021}. 

The wavelength agreement is the primary reason why some observed lines cannot be ranked as ``A'', and the uncertain identifications done manually could be attributed as another cause.  For example, in the MagE spectrum of J0608, an extremely weak absorption feature at 3152.75\,{\AA} was originally identified as \ion{Fe}{i} $\lambda$3152.01, even with the condition that an obvious wavelength disagreement and a low transition probability both exist.  Another example is the absorption line at 3735.16\,{\AA}, which was identified as \ion{H}{i} $\lambda$3734.37 in \citet{Williams2021}.  PyEMILI suggests that \ion{Ne}{ii} $\lambda$3734.94 is the ID with the highest level of confidence, because the other four fine-structure lines of \ion{Ne}{ii} belonging to the same multiplet have been detected in the spectrum and a smaller wavelength difference with respect to the observed line compared to \ion{H}{i} $\lambda$3734.37.  Besides, other \ion{H}{i} lines detected in the spectrum are all emission lines.  Thus the most confident ID of this absorption feature at 3735.16\,{\AA} should be \ion{Ne}{ii} rather than \ion{H}{i}.  Complete output of PyEMILI's identification for tJ0608 can be found on the webpage (see {\sc url} given in Section\,\ref{pyemili:sec1}).

\section{Summary and Discussion} 
\label{sec4}

A new generation {\sc Python}-based emission line identification code PyEMILI has been developed in the context that deep, high-dispersion spectra of various emission-line objects are readily to obtain in the era of data-intensive astronomy today, when reliable, massive identification of the rich emission-line spectra with high efficiency is urgently needed.  We have designed a complete logic for PyEMILI in the procedure of spectral-line identifications, wherein three basic criteria of line identification were used.  The three criteria, although basically same as those used by EMILI, have been further refined and significantly upgraded for PyEMILI so that the algorithm of its spectral-line identification is more suitable for modern astronomical spectroscopy.  Meanwhile, we provided a brief overview of spectral line formation. 

For a comprehensive identification of spectral lines, a complete (although still difficult to achieve so far) atomic transition database is indispensable.  For the atomic transition database used by PyEMILI, we have expanded the volume of the Atomic Line List v3.00b4 of \citet{2018vanhoof} by adding a large number of transition probabilities given in the literature, including the Kurucz Line Lists.  In order to help identify the numerous optical recombination lines, the majority of which are very faint, in the deep high-dispersion spectra of photoionized gaseous nebulae, we have compiled for PyEMILI a new dataset of effective recombination coefficients for nebular emission lines, particularly those of the relatively abundant heavy-element ions (\ion{C}{ii}, \ion{N}{ii}, \ion{O}{ii}, and \ion{Ne}{ii}) present in PNe and \ion{H}{ii} regions, from the literature. 

We tested PyEMILI by running the code to identify the emission lines as measured from the deep spectroscopic data of three representative emission-line objects, Galactic PNe IC\,418 and Hf\,2-2 and a late-type [WC11] Wolf-Rayet star J0608 in the LMC, all of which have the published line lists with manual identifications that can be used for purpose of comparison.  The results of PyEMILI's line identifications for the three objects all show improvements over previous manual work. 

\smallskip

There is no absolute benchmark of line identification in the whole optical wavelength region (from $\sim$3100\,{\AA} to 1\,$\mu$m, as covered by the majority of the ground-based facilities), especially for those rare weak lines, when the elemental distribution and ionic abundances of the emission-line object being investigated are unclear.  The algorithm of PyEMILI in line identification is largely based on the traditional method.  Hence a high level of consistency between the results of PyEMILI's identification and those of the manual identifications does not necessarily mean a high confidence level of correctness in line identifications. 

Reliable line identifications are largely based on accurate PyEMILI model of the object being investigated (Section\,\ref{pyemili:sec2}), which in turn depends on correct identification of the nebular lines that can be used to refine the models (as demonstrated in Figure\,\ref{flowchart}).  The identification of faint emission lines in a nebular spectrum is thus based on the nebular models defined by the strong lines, whose transition, excitation and formation are already well understood.  However, in the nebular spectrum there are weak emission lines formed by mechanisms other than collisional excitation and recombination, such as fluorescence and charge exchange; emission lines formed by these mechanisms do not always follow the energy-bin model as described in Section\,\ref{pyemili:sec2}.  Therefore, there could still be underlying uncertainty in the identification of some of the faint lines, either manually or by machine, unless a large sample of objects with deep high-dispersion spectroscopy has been studied in order to reach consensus on the correct identification of these faintest lines.  What PyEMILI can do is to provide an observed spectral line with the most probable candidate ID(s) using the well refined criteria of identification and a general and reasonable model. 

We have tried to use a series of electron temperature and density combinations to iteratively identify the input lines till convergence on an optimal electron temperature and density was reached.  However, due to uncertainty in the identification of very faint lines, it is difficult to define a better direction of iterative runs as well as optimal results of line identification.  Furthermore, the adopted assumption of a uniform nebular electron temperature ignores temperature fluctuation in nebulae.  We therefore give up on finding the optimal electron temperature and density, and instead, resort to the expectation that reasonable results could be obtained with default temperature and density.  We thus did not utilize the Saha ionization equation to derive the ionic abundances. 
In the case of absorption lines, we should still be careful with the electron temperature as the population of the lower energy level is sensitive to the temperature. 

The atomic transition database used by PyEMILI, although already significantly expanded (see description in Section\,\ref{pyemili:sec3}), might not cover all transitions from 1000\,{\AA} to 2\,$\mu$m.  Several transitions of common ionic species, mostly very faint, may still be missing from the database; this was found when we tested the deep spectra of Hf\,2-2.  Several \ion{O}{ii} lines of intermediate coupling, in particular those belonging to the 3d--4f transition array, e.g.\ $\lambda$4291.25, $\lambda$4344.37 and $\lambda$4477.91 coming from the upper spectral term G[3]$^{\rm o}$, were not found in the current atomic transition database.  We have been supplementing the database using various sources.  The atomic transition database with newly added atomic transitions will be updated online timely (see Section\,\ref{sec5}).

\section{Accessibility of the Code} 
\label{sec5}

The source code of PyEMILI, as well as all the related data (the input line lists, the original outputs of PyEMILI runs, and the final identified line lists of the three test objects, IC\,418, Hf\,2-2 and J0608) is available on Zenodo under an open-source Creative Commons Attribution 4.0 International License (CC-BY): \underline{\dataset[doi:10.5281/zenodo.14054096]{https://doi.org/10.5281/zenodo.14054096}} (permanent DOI).

The PyEMILI code, user manual (still in revision), and related data have also been archived on GitHub\footnote{\url{https://github.com/LuShenJ/PyEMILI};  this link has been connected to the Zenodo repository (\dataset[doi:10.5281/zenodo.14054096]{https://doi.org/10.5281/zenodo.14054096}), so that any update of the material on GitHub can be synchronized timely on Zenodo.} and already publicly available.  The source code of PyEMILI can now be downloaded directly and freely from GitHub; installation of the code on local machines can be completed by compiling the file {\sc setup.py} using Python\,3. 

The PyEMILI code will soon be publicly available online for free installation via the Python Package Index (PyPI\footnote{\url{https://pypi.org}}).  We have successfully tested installation of the code from PyPI using the {\sc Python} package installer {\sc pip} (v24.3.1).

\section{Further Improvements and Future Prospects} 
\label{sec6}

We will make follow-up improvements in PyEMILI, and have designed a long-term plan to further enhance the capability of the code in identifying spectral lines from far-UV to far-IR.

\subsection{Nebular Analysis with Heavy-element ORLs}
\label{sec6:part1}

In order to facilitate reliable identification of faint nebular emission lines, in particular the numerous ORLs of \ion{C}{ii}, \ion{N}{ii}, \ion{O}{ii} and \ion{Ne}{ii} detected in the deep, high-dispersion spectra of PNe and \ion{H}{ii} regions, we have compiled a dataset of the effective recombination coefficients of these nebular lines (see Section\,\ref{ERC}) and incorporated it into the atomic transition database used by PyEMILI.  These coefficients are used to calculate the predicted line fluxes via Eq\,\ref{pred_flux}. 

The predicted line flux of an ORL, $I_{\rm pred}$($\lambda$), is a function of electron temperature and electron density.  At a given $T_{\rm e}$ and $N_{\rm e}$ and using an assumed ionic abundance (X$^{2+}$/H$^{+}$), we can use $I_{\rm pred}$($\lambda$) of numerous transitions to create a theoretical spectrum with a given spectral resolution $R$.  We will create a subroutine in the PyEMILI code to read the dataset of effective recombination coefficients and generate the theoretical spectra of heavy-element ORLs at various temperature and density cases.  These theoretical recombination spectra can be used to fit the observed spectrum of a PN (or an \ion{H}{ii} region), to obtain the electron temperature, electron density and ionic abundances of the target nebula using the Markov-Chain Monte Carlo (MCMC) method.  This follow-up improvement in PyEMILI, including test runs of the code on a carefully selected sample of 34 emission-line objects with deep high-dispersion spectroscopic data collected from the literature, will sbe reported in a separate paper (Z.\ Tu et al., in preparation).

\subsection{Identification of Mid-IR Spectral Lines}
\label{sec6:part2}

The current version of PyEMILI was designed to identify spectral lines from $\sim$1000\,{\AA} to 2\,$\mu$m, a wavelength region covered by the majority of the high-dispersion spectrographs on the ground-based large telescopes (e.g., $\sim$3100--10,000\,{\AA} covered by Subaru/HDS, $\sim$3000--25,000\,{\AA} covered by VLT/XSHOOTER, and 3000--10,000\,{\AA} covered by Keck/HIRES; \citealt{1994SPIE.2198..362V,Vernet_2011,2014AN....335...27A}) as well as the \emph{Hubble Space Telescope} (\emph{HST}, 1150--10,300\,{\AA} covered by the Space Telescope Imaging Spectrograph, STIS onboard).  However, with the launching and science operation of the \emph{James Webb Space Telescope} (\emph{JWST}), an increasing number of high-quality mid-IR spectra are becoming available \citep[e.g.][]{Wesson_2024}.  It is thus imperative to add the capability of the PyEMILI code in identifying spectral lines in the mid-IR wavelengths up to $\sim$28\,$\mu$m, to accommodate the wavelength coverage of the Near-Infrared Spectrograph (NIRSpec, 0.6--5.3\,$\mu$m; \citealt{Jakobsen_2022,Boker_2023}) and the Mid-Infrared Instrument (MIRI, 4.9--27.9\,$\mu$m; \citealt{2023PASP..135d8003W,2023A&A...675A.111A}) on the \emph{JWST}. 

Following the release of the current version of PyEMILI (Section\,\ref{sec5}), we have started further improvement of the code so that it will be applicable to the mid-IR spectra (2--28\,$\mu$m).  This task not only involves expansion of the atomic transition database by including transitions in 2--28\,$\mu$m, but also involves a thorough revision of the code (including re-assessment of the major criteria for line identification, as well as various approximations and optimizations used in the source code) accordingly; subsequently test runs of the code using high-quality \emph{JWST} spectra, which are still in limited quantity, are also needed.  This further improvement is very time-consuming and requires tremendous efforts, and thus worth a separate project.

\subsection{Future Development and Prospects}
\label{sec6:part3}  

It is worth mentioning that the selection of the weight for each of the three criteria of PyEMILI in spectral-line identification, although delicately designed and carefully tested, are still not all perfect.  As more and more manual identifications of spectral lines with high confidence levels become available, thanks to the availability of deep high-dispersion spectroscopy of emission-line objects with large telescopes, the line identification criteria utilized by PyEMILI as well as the architecture of the code design may be further optimized in the future.  The current line identification logic of PyEMILI is more comprehensive than its predecessor EMILI, with more interactive measures being implemented, such as a cross-check of the A-ranking candidates within the same multiplet, including the cross-check of transition series (such as the Balmer series of \ion{H}{i} and series of \ion{He}{ii} lines) with the same lower level but different upper levels. 

The current version of PyEMILI can run more automatically and provides satisfying results based on its internal general models.  A comparison of the identification results of the emission-lines in the Galactic PN IC\,418 shows that PyEMILI can produce a higher rate of agreement than EMILI does.  The identification results of the late-type [WC11] star J0608 in LMC also exhibit the high extensibility of PyEMILI.  In our future plan, we will make PyEMILI fully applicable to all spectra, including emission-line and absorption-line. 

As of 2024 November, there are over 300 kinds of molecules that have been detected in the interstellar medium and circumstellar envelopes\footnote{\url{https://cdms.astro.uni-koeln.de/classic/molecules}}.  Emission of these molecules spans from UV, optical, IR to radio.  There are also a number of IR emission features that may come from unknown species/molecules and are considered as unidentified emission (UIE).  Moreover, although still in very limited quantity, high-quality near- to mid-IR spectra of PNe obtained with \emph{JWST} are becoming available, and numerous molecular emission lines were detected \citep{2022NatAs...6.1421D,Wesson_2024}.  Future development of PyEMILI will also involve the identification of molecular lines from optical to far-IR; with this capability, PyEMILI can be used to identify ro-vibrational transitions of H$_{2}$, CO, CH, OH, fullerenes, polycyclic aliphatic hydrocarbon (PAH), various interstellar ices, etc. (Y.\ Zhang 2024, private communication).  This further development (PyEMILI$+$, as might be named) requires even greater efforts, including not only the compilation of a huge volume of theoretical ro-vibrational transition database of molecules (given the complexity of molecular structures), but probably also a new design of the spectral-line identification criteria. 

Our ultimate goal is to eventually develop PyEMILI into a more standard, general-purpose package that comprises an extensive database of atomic and molecular transitions for spectral-line identification, and make it applicable to almost all types of spectra (including emission-line and absorption-line) of various celestial objects, including emission-line nebulae and stellar objects.  This endeavour will have a profound impact on astronomical spectroscopy with high S/N ratios in the modern era.

\begin{deluxetable*}{ccccccccccccc}
\setlength{\tabcolsep}{0.3cm}
\tablecaption{Line List of Hf\,2-2 \label{hf2-2_linelist}}
\tablehead{\colhead{$\lambda_{\mathrm{obs}}$} & \colhead{$F_\lambda/F_{\mathrm{H\beta}}$} & \colhead{$I_\lambda/I_{\mathrm{H\beta}}$} &  \colhead{Error} & \colhead{$IDI$} & \colhead{ID} & \colhead{$\lambda_{\mathrm{lab}}$} & \colhead{Transition} & \colhead{Electron Configurations} & \colhead{$g_\mathrm{l}$} & \colhead{$g_\mathrm{u}$} \\ 
\colhead{(\AA)} & \colhead{} & \colhead{} & \colhead{(\%)} & \colhead{} & \colhead{} & \colhead{(\AA)} & \colhead{(Lower-Upper)} & \colhead{(Lower-Upper)} & \colhead{} & \colhead{} } 
\startdata
3132.84 & 2.10E-02 & 3.47E-02 & 2.45 & 1A & \ion{O}{iii} & 3132.79 & $\mathrm{^3S\text{--}{^3P^o}}$ & $\mathrm{2s^2.2p.({^2P^o}).3p\text{--}2s^2.2p.({^2P^o}).3d}$ & 3 & 5 \\
3187.78 & 4.18E-02 & 6.72E-02 & 2.31 & 2A & \ion{He}{i} & 3187.74 & $\mathrm{^3S\text{--}{^3P^o}}$ & $\mathrm{1s.2s\text{--}1s.4p}$ & 3 & 1 \\
3203.15 & 1.40E-02 & 2.23E-02 & 3.77 & 1A & \ion{He}{ii} & 3203.10 & $\mathrm{3\text{--}5}$ & $\mathrm{3^*\text{--}5^*}$ & 6 & 10 \\
3218.23 & 7.83E-03 & 1.24E-02 & 4.63 & 2A & \ion{Ne}{ii} & 3218.19 & $\mathrm{^4D^o\text{--}{^4F}}$ & $\mathrm{2s^2.2p^4.({^3P}).3p\text{--}2s^2.2p^4.({^3P}).3d}$ & 8 & 10 \\
3244.05 & 4.90E-03 & 7.68E-03 & 8.43 & 2A & \ion{Ne}{ii} & 3244.10 & $\mathrm{^4D^o\text{--}{^4F}}$ & $\mathrm{2s^2.2p^4.({^3P}).3p\text{--}2s^2.2p^4.({^3P}).3d}$ & 6 & 8 \\
3258.26 & 2.62E-03 & 4.08E-03 & 13.92 & 1A & \ion{He}{i} & 3258.27 & $\mathrm{^1S\text{--}{^1P^o}}$ & $\mathrm{1s.2s\text{--}1s.9p}$ & 1 & 3 \\
3271.68 & 2.40E-03 & 3.71E-03 & 13.67 & 2A & \ion{N}{ii} & 3271.62 & $\mathrm{^3F^o\text{--}{^3D}}$ & $\mathrm{2s^2.2p.({^2P^o}).3d\text{--}2s^2.2p.({^2P^o}).5p}$ & 9 & 7 \\
3312.40 & 2.28E-03 & 3.48E-03 & 11.52 & 2A & \ion{O}{iii} & 3312.33 & $\mathrm{^3P^o\text{--}{^3S}}$ & $\mathrm{2s^2.2p.({^2P^o}).3s\text{--}2s^2.2p.({^2P^o}).3p}$ & 3 & 3 \\
3334.87 & 7.05E-03 & 1.07E-02 & 3.56 & 2A & \ion{Ne}{ii} & 3334.84 & $\mathrm{^4P\text{--}{^4D^o}}$ & $\mathrm{2s^2.2p^4.({^3P}).3s\text{--}2s^2.2p^4.({^3P}).3p}$ & 6 & 8 \\
3340.82 & 2.92E-03 & 4.42E-03 & 8.79 & 2A & \ion{O}{iii} & 3340.76 & $\mathrm{^3P^o\text{--}{^3S}}$ & $\mathrm{2s^2.2p.({^2P^o}).3s\text{--}2s^2.2p.({^2P^o}).3p}$ & 5 & 3 \\
3354.98 & 5.80E-03 & 8.72E-03 & 3.81 & 2A & \ion{Ne}{ii} & 3355.02 & $\mathrm{^4P\text{--}{^4D^o}}$ & $\mathrm{2s^2.2p^4.({^3P}).3s\text{--}2s^2.2p^4.({^3P}).3p}$ & 4 & 6 \\
\enddata
\tablecomments{The observed wavelengths were corrected for a radial velocity $V_{\mathrm{ral}}=44.9\,\mathrm{km\,s^{-1}}$. Error: Observed line flux error. IDI: PyEMILI IDI value/ranking.\\
(This table is published in its entirety online only in the machine-readable format.)}
\end{deluxetable*}

\begin{deluxetable*}{ccccccccc}
\setlength{\tabcolsep}{0.45cm}
\tablecaption{Line List of J0608\label{j0608linelist}}
\tablehead{\colhead{$\lambda_{\mathrm{obs}}$$^{a}$} & \colhead{$F_\lambda/F_{\mathrm{H\beta}}$$^{b}$} & \colhead{$IDI$} & \colhead{ID$^{c}$} & \colhead{$\lambda_{\mathrm{lab}}$} & \colhead{Transition} & \colhead{Electron Configurations} & \colhead{$g_\mathrm{l}$} & \colhead{$g_\mathrm{u}$} \\ 
\colhead{(\AA)} & \colhead{} & \colhead{} & \colhead{} & \colhead{(\AA)} & \colhead{(Lower-Upper)} & \colhead{(Lower-Upper)} & \colhead{} & \colhead{} } 
\startdata
3134.82 &   $-$1.50E-01 & 1A & \ion{O}{ii} & 3134.73 & $\mathrm{{^4}D^o\text{--}{^4}P}$ & $\mathrm{2s^2.2p^2.({^3}P).3p\text{--}2s^2.2p^2.({^3}P).4s}$ & 8 & 6 \\
3139.55 &   $-$2.27E-02 & 2A & \ion{O}{ii} & 3139.68 & $\mathrm{{^4}D^o\text{--}{^4}P}$ & $\mathrm{2s^2.2p^2.({^3}P).3p\text{--}2s^2.2p^2.({^3}P).4s}$ & 4 & 2 \\
3152.75 &   $-$1.96E-02 & 6B & \ion{Fe}{i} & 3152.01 & $\mathrm{z{^5}F^o\text{--}{^5}F}$ & $\mathrm{3d^6.({^5}D).4s.4p.({^3}P^o)\text{--}3d^7.({^4}F).5d}$ & 11 & 11 \\
3165.54 &      2.65E+00 & 1A & \ion{C}{ii} & 3165.46 & $\mathrm{{^2}D^o\text{--}{^2}P}$ & $\mathrm{2p^3\text{--}2s.2p.({^3}P^o).3p}$ & 6 & 4 \\
3167.87 &      1.82E+00 & 1A & \ion{C}{ii} & 3167.94 & $\mathrm{{^2}D^o\text{--}{^2}P}$ & $\mathrm{2p^3\text{--}2s.2p.({^3}P^o).3p}$ & 4 & 2 \\
3187.88 &      6.52E-01 & 2A & \ion{He}{i} & 3187.74 & $\mathrm{{^3}S\text{--}{^3}P^o}$ & $\mathrm{1s.2s\text{--}1s.4p}$ & 3 & 1 \\
3214.21 &   $-$1.43E-01 & 2A & \ion{Ne}{ii} & 3214.33 & $\mathrm{{^4}D^o\text{--}{^4}F}$ & $\mathrm{2s^2.2p^4.({^3}P).3p\text{--}2s^2.2p^4.({^3}P).3d}$ & 4 & 6 \\
3218.20 &   $-$3.24E-01 & 2A & \ion{Ne}{ii} & 3218.19 & $\mathrm{{^4}D^o\text{--}{^4}F}$ & $\mathrm{2s^2.2p^4.({^3}P).3p\text{--}2s^2.2p^4.({^3}P).3d}$ & 8 & 10 \\
        &               & 2A & \ion{O}{ii} & 3217.93 & $\mathrm{{^6}P\text{--}{^6}S^o}$ & $\mathrm{2s.2p^3.({^5}S^o).3p\text{--}2s.2p^3.({^5}S^o).4s}$ & 8 & 6 \\
3230.10 &   $-$2.33E-01 & 1A & \ion{Ne}{ii} & 3230.07 & $\mathrm{{^2}D\text{--}{^2}D^o}$ & $\mathrm{2s^2.2p^4.({^1}D).3s\text{--}2s^2.2p^4.({^1}D).3p}$ & 6 & 6 \\
3232.54 &   $-$5.12E-02 & 1A & \ion{Ne}{ii} & 3232.37 & $\mathrm{{^2}D\text{--}{^2}D^o}$ & $\mathrm{2s^2.2p^4.({^1}D).3s\text{--}2s^2.2p^4.({^1}D).3p}$ & 4 & 4 \\
3279.01 &      4.32E-01 & ... & ... & ... & ... & ... & ... & ... \\
3297.98 &   $-$1.82E-01 & 0A & \ion{Ne}{ii} & 3297.73 & $\mathrm{{^4}P\text{--}{^4}D^o}$ & $\mathrm{2s^2.2p^4.({^3}P).3s\text{--}2s^2.2p^4.({^3}P).3p}$ & 6 & 6 \\
\enddata
\tablecomments{ 
\tablenotetext{a}{The $\lambda_{\mathrm{obs}}$ have been corrected radial velocities $V_{\mathrm{ral}}=232.1\,\mathrm{km\,s^{-1}}$ for emission lines and $V_{\mathrm{ral}}=211.6\,\mathrm{km\,s^{-1}}$ for absorption lines measured by \citet{2020ApJ...888...54M}.}
\tablenotetext{b}{A negative sign means that this is an absorption line, and the value following the negative sign represents the equivalent width (in the units of \AA) of that absorption line.}
\tablenotetext{c}{These IDs are adopted from \citet{Williams2021}.}
(This table is published in its entirety online only in the machine-readable format.)}
\end{deluxetable*}

\begin{acknowledgments}
We are grateful to the anonymous referee whose excellent comments and suggestions greatly improved this article.  We thank Dr. Peter A.~M. van Hoof (Royal Observatory of Belgium, Brussels), developer of the Atomic Line List v3.00b4, for in-depth discussion and suggestions during our development and revision of the PyEMILI code.  We also thank Dr.\ Yong Zhang (School of Physics and Astronomy, Sun Yat-Sen University, Zhuhai) for discussion and suggestions.  X.F.\ acknowledges support from the Youth Talent Program (2021) from the Chinese Academy of Sciences (CAS, Beijing), and the ``Tianchi Talents'' Program (2023) of the Xinjiang Autonomous Region, China. 
This work is supported by the National Natural Science Foundation of China (NSFC) through project 11988101. 
\end{acknowledgments}

\section*{Data availability}
Tables\,\ref{tab:exam_ic418}, \ref{hf2-2_linelist} and \ref{j0608linelist} are published in their entirety online in the machine-readable format, and are also available in electronic forms at CDS.


\smallskip

\software{NumPy \citep{numpy},
          SciPy \citep{scipy},
          Matplotlib \citep{matplotlib},
          Astropy \citep{astropy},
          {pandas} \citep{pandas},
          Numba \citep{numba}
          }

\bibliography{PyEMILI}{}
\bibliographystyle{aasjournal}

\end{document}